\documentclass[aps,prb,superscriptaddress,citeautoscript,reprint,twocolumn]{revtex4-2}

\usepackage[english]{babel} 
\usepackage[utf8]{inputenc} 
\usepackage[T1]{fontenc} 

\usepackage{graphicx} 

\usepackage{fancyhdr} 
\usepackage[colorlinks=true,bookmarks=false,citecolor=blue,linkcolor=red,urlcolor=blue]{hyperref}

\usepackage{amsthm,amsmath,amssymb,amsfonts,bbm}
\usepackage{array} 
\usepackage{gensymb} 
\usepackage[caption=false]{subfig} 
\usepackage{lpic}
\usepackage{color}
\usepackage{transparent}
\usepackage{listings} 
\usepackage{booktabs}
\usepackage{multirow}
\usepackage[dvipsnames]{xcolor}
\usepackage{ulem}

\providecommand{\e}[1]{\ensuremath{\times 10^{#1}}}

\begin{document}

\title{Observing Magnetic Monopoles in Spin Ice using Electron Holography}

\author{Ankur Dhar}
\email{ankur.dhar@alumni.oist.jp}
\affiliation{Quantum Wave Microscopy Unit, Okinawa Institute of Science and Technology}
\affiliation{Theory of Quantum Matter Unit, Okinawa Institute of Science and Technology}
\author{L.\ D.\ C.\ Jaubert}
\email{ludovic.jaubert@cnrs.fr}
\affiliation{Theory of Quantum Matter Unit, Okinawa Institute of Science and Technology}
\affiliation{CNRS, Universit\'e de Bordeaux, LOMA, UMR 5798, 33400 Talence, France}
\author{Cathal Cassidy}
\email{c.cassidy@oist.jp}
\affiliation{Quantum Wave Microscopy Unit, Okinawa Institute of Science and Technology}
\author{Tsumoru Shintake}
\email{shintake@oist.jp}
\affiliation{Quantum Wave Microscopy Unit, Okinawa Institute of Science and Technology}
\author{Nic Shannon}
\email{nic.shannon@oist.jp}
\affiliation{Theory of Quantum Matter Unit, Okinawa Institute of Science and Technology}

\date{\today}

\begin{abstract}

While there is compelling evidence for the existence of magnetic monopoles in spin ice, 
the direct observation of a point-like source of magnetic field in these systems remains 
an open challenge. 
One promising approach is electron holography, which combines atomic-scale resolution 
with extreme sensitivity to magnetic vector potentials, through the Aharonov-Bohm effect.     
Here we explore what holography can teach us about magnetic monopoles in spin ice, 
through 
experiments on artificial spin ice, and numerical simulations of pyrochlore spin ice.
In the case of artificial spin ice, we show that holograms can be used to measure local 
magnetic charge. 
For pyrochlore spin ice, we demonstrate that holographic experiments are capable 
of resolving both magnetic monopoles and their dynamics, including the emergence 
of electric fields associated with fluctuations of closed loops of spins. 
These results establish that the observation of both magnetic monopoles and emergent 
electric fields in pyrochlore spin ice is a realistic possibility in an electron microscope with 
sufficiently high phase resolution.

\end{abstract}

\maketitle

\begin{figure*}
	\centering
	\subfloat[]{\includegraphics[width=.3\textwidth]{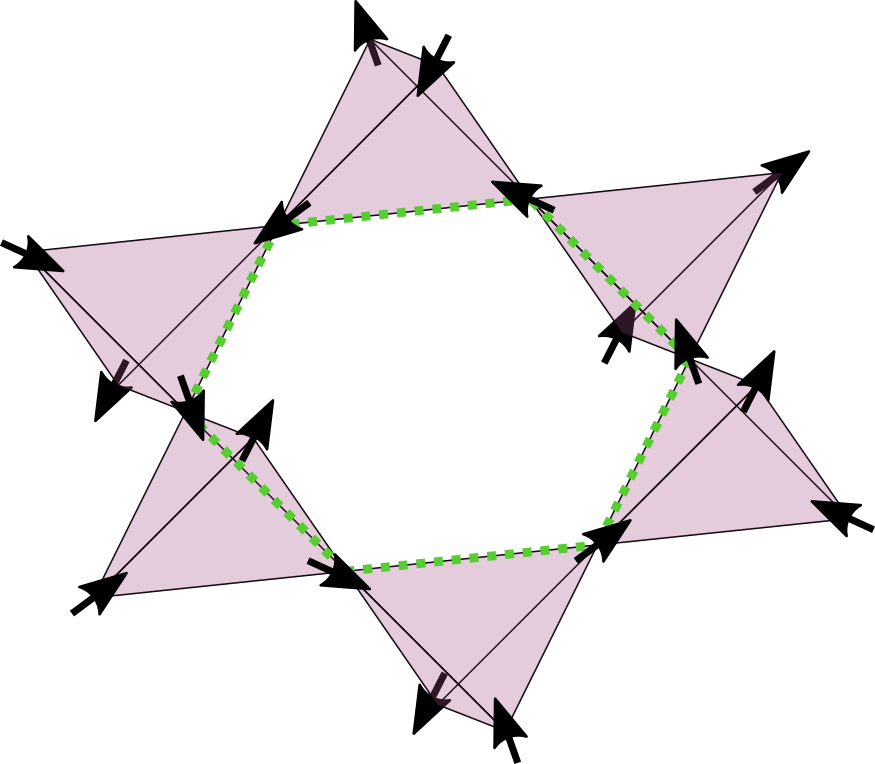}\label{fig:ice.rules}}
	\subfloat[]{\includegraphics[width=.3\textwidth]{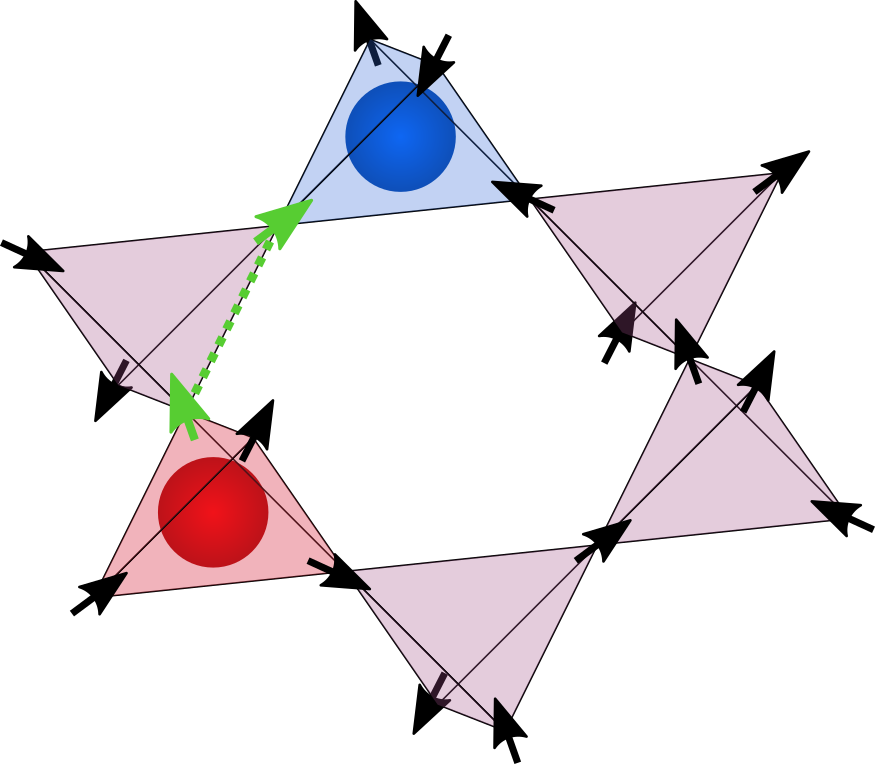}\label{fig:monopoles}}
	\subfloat[]{\includegraphics[width=.3\textwidth]{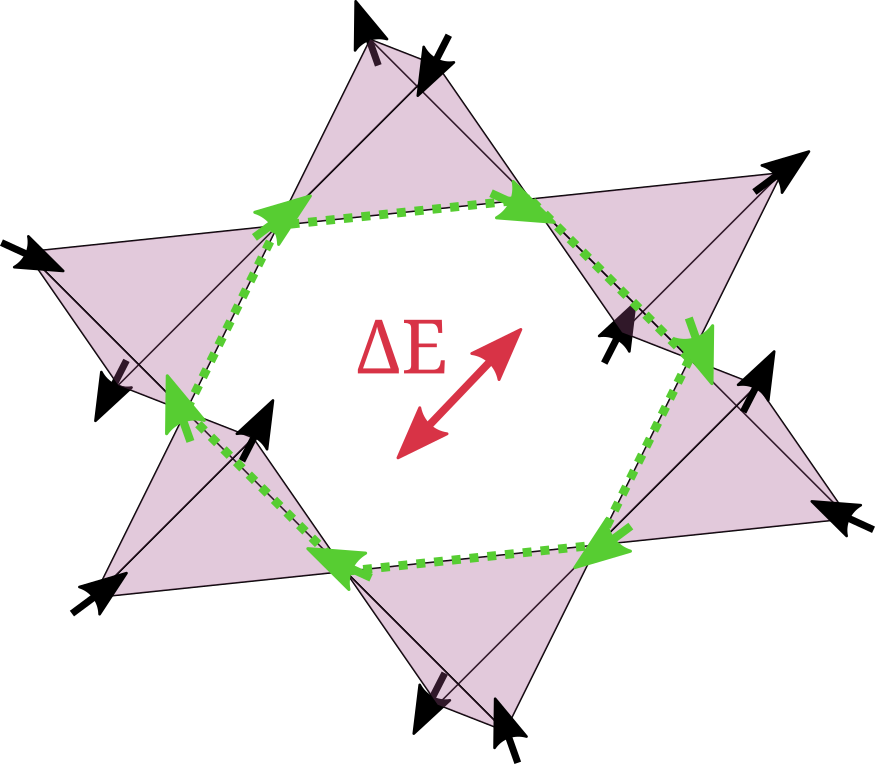}\label{fig:electric.field}}
	\caption{\textbf{Spin ice, its emergent magnetic monopoles and 
	electric field.} 
	\textbf{(a)}  
	Pyrochlore lattice in spin--ice materials, formed of corner--sharing tetrahedra.  
	The spin configuration shown obeys the ``ice rules'', in which two spins 
	point out of, and two spins point into, each tetrahedron in the lattice.
	A ``flippable'' plaquette, in which spins point head--to-tail in a closed loop, 
	is highlighted in green.
	\textbf{(b)}  Flipping a single spin causes two neighbouring tetrahedra to violate 
	the ice rules, 
	and act as sources and sinks of magnetic field (magnetic monopoles).
	These monopole excitations (red/blue) can then move within 
	the lattice by successive spin flips (green arrows). 
	\textbf{(c)} 
	A pair of monopoles can recombine by traveling around a 
	flippable loop of spins (green arrows).
	Just as electric current traveling in a loop acts as a source of magnetic field, 
	so the transport of magnetic charge around this hexagonal plaquette induces 
	an electric field.}
	\label{fig:spin.ice}
\end{figure*}

Although the concept of magnetic monopoles --- particles which act as point--sources 
of magnetic field --- is over a century old~\cite{Curie1894,Poincare1896,Dirac1931},
their existence remains an enigma \cite{Cabera1982}.  
While Dirac conceived of magnetic monopoles as elementary particles in 
a vacuum \cite{Dirac1931}, the best--characterised examples arise as 
monopoles of the magnetic field (${\bf H}$) within the pyrochlore magnets 
known as ``spin ice'' \cite{Castelnovo2008} [Fig.~\ref{fig:spin.ice}], 
and their artificial cousins \cite{Wang2006}.   
Great ingenuity has been brought to the detection of monopoles in spin ice 
\cite{Jaubert2009,Bramwell2009,Blundell2012,Morris2009,Dusad2019,Samarakoon21a}
exploiting analogies with 
electrolytes \cite{Castelnovo2008,Jaubert2009,Bramwell2009,Blundell2012}; 
Dirac strings \cite{Morris2009}; 
and even the ``sound'' made by monopole motion \cite{Dusad2019,Samarakoon21a}.
None the less, direct observation of an individual monopole remains 
an open challenge.


The method we pursue is electron holography.
An electron wavefront incident on a point--source of magnetic field 
undergoes a profound change, acquiring a phase which winds 
around its axis, so the wavefunction describes a ``vortex'' with 
finite angular momentum \cite{Tamm1931,Fiertz1944,Wilczek1982}.   
The ideal technique for measuring this phase is holography, in which an image 
is created through the interference of coherent waves~\cite{Gabor1948}.   
In recent years, the use of electron holography to image microscopic magnetic structures 
has been raised to a high art \cite{Lichte2002,McCartney2005,Fert2013,Park2014,Zheng2017}, 
and proof--of--principle holographic measurements of a magnetic needle, as a macroscopic 
monopole analogue, have already been reported \cite{Beche2014}.
In this Article, we extend electron holography to the emergent monopoles of spin ice, 
with the goal of directly imaging both their magnetic charge, and dynamics.
We do this through experimental measurements of artificial spin ice, and detailed 
simulations of holographic measurements of pyrochlore spin ice.


Our experiments on artificial spin ice confirm that holography can be used to 
characterise the magnetic monopoles of a spin--ice like system, 
providing quantitative measurements of their (quantized) magnetic charge.
Meanwhile, simulations of 
thin films of pyrochlore spin ice show how holography could be used both 
to image magnetic monopoles, and to study their dynamics.
We exhibit the characteristic phase 
map associated with monopoles, and establish that electron holography could also 
be used to resolve the emergent electric field found in a spin ice with dynamics.  
We also provide estimates of the instrumental requirements 
needed to resolve these emergent excitations.
These results establish that the observation of both magnetic monopoles 
and emergent electric fields in pyrochlore spin ice is a realistic possibility,  
in an electron microscope with sufficiently--high phase resolution.


Spin ice, exemplified by ${\mathrm{Dy}}_{2}{\mathrm{Ti}}_{2}\mathrm{O}_{7}$ 
and ${\mathrm{Ho}}_{2}{\mathrm{Ti}}_{2}\mathrm{O}_{7}$~\cite{Harris1997}, is a family 
of magnetic insulators with a pyrochlore lattice [Fig.~\ref{fig:spin.ice}]. 
Magnetic ions have Ising moments, and at low temperatures obey 
the ice--rules \cite{Bernal1933,Pauling1935}, with exactly two moments (spins) 
pointing into and two out of, each tetrahedron in the lattice, such that 
the divergence of the magnetization 
$\nabla \cdot {\bf M} = - \nabla \cdot {\bf H} = 0$ [Fig.~\ref{fig:ice.rules}]. 
Flipping a single spin creates a source and sink of magnetic field in 
neighbouring tetrahedra, $\nabla \cdot {\bf H} = \pm q_m$ 
\cite{Ryzhkin2005,Castelnovo2008}, and successive spin--flips permit both 
excitations to move, independently, through the lattice  \cite{Jaubert2009}.
These are the magnetic monopoles of spin ice [Fig.~\ref{fig:monopoles}], 
which share many of the properties of Dirac's monopoles~\cite{Dirac1931}, 
including Coulomb interactions, and strict quantization of 
magnetic charge~\cite{Castelnovo2008}.
Where magnetic moments behave like idealised magnet dipoles, 
this charge can be calculated through the ``dumbbell model''~\cite{Castelnovo2008}, 
and for Dy$_2$Ti$_2$O$_7$ is given by
\begin{eqnarray} 
	q_m^{\rm SI} = 2m/d = 4.3\times 10^{-13}\ \text{A.m} \; , 
\label{eq:qm.spin.ice}
\end{eqnarray} 
where $m = 10 \mu_B$ is the moment, and $d = \sqrt{3}/4\ \textrm{nm}$ 
is the distance between the centers of neighbouring tetrahedra.
While there is compelling evidence for the existence of monopoles in spin ice
\cite{Jaubert2009,Morris2009,Bramwell2009,Blundell2012,Dusad2019,Samarakoon21a}, 
their direct observation 
remains a significant challenge, demanding sensitivity to magnetic fields 
on the scale of a single tetrahedron.


\begin{figure*}
	\centering
	\subfloat[]{\includegraphics[height=.2\textheight]{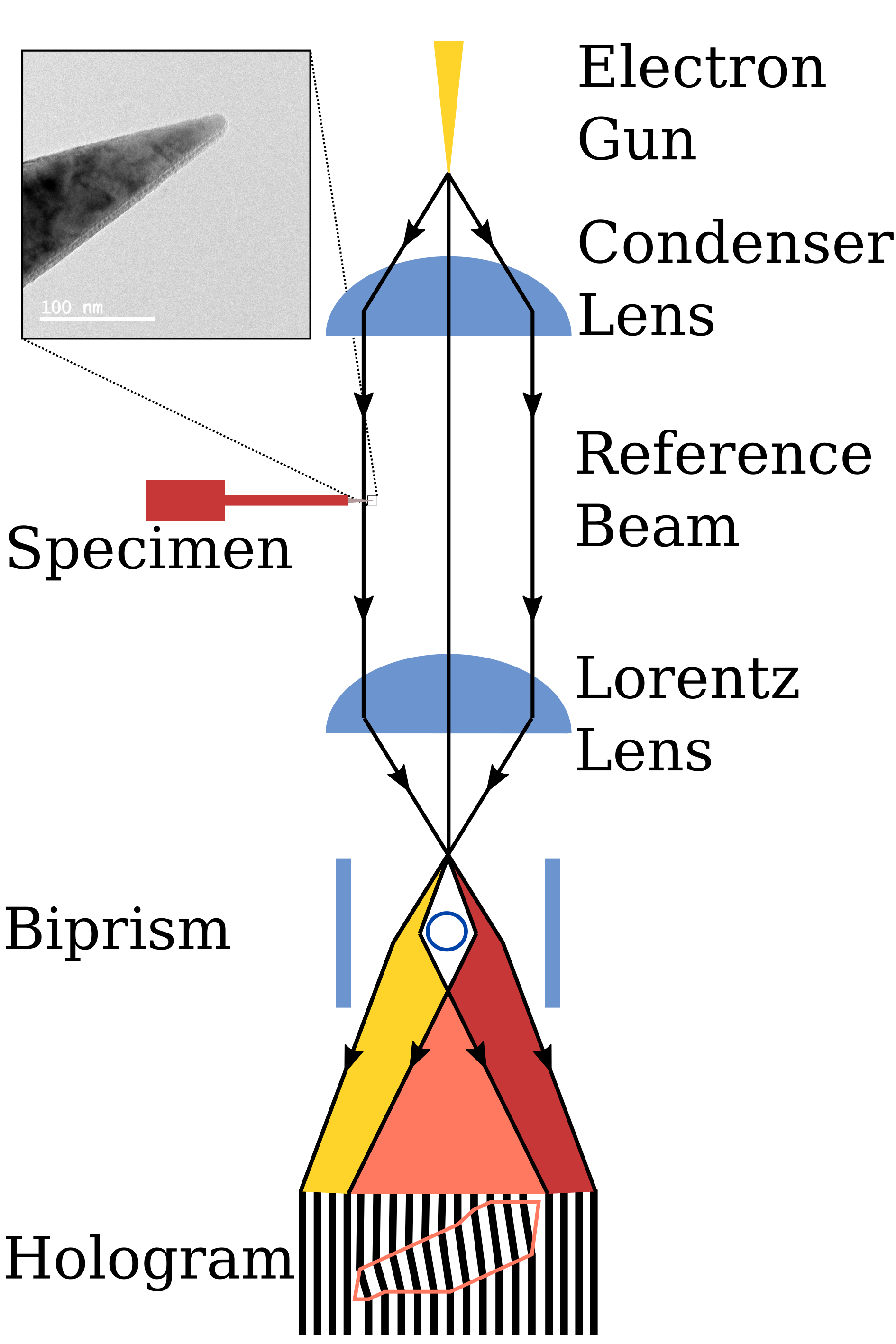}\label{fig:microscope}}\hfil
	\subfloat[]{\includegraphics[height=.2\textheight]{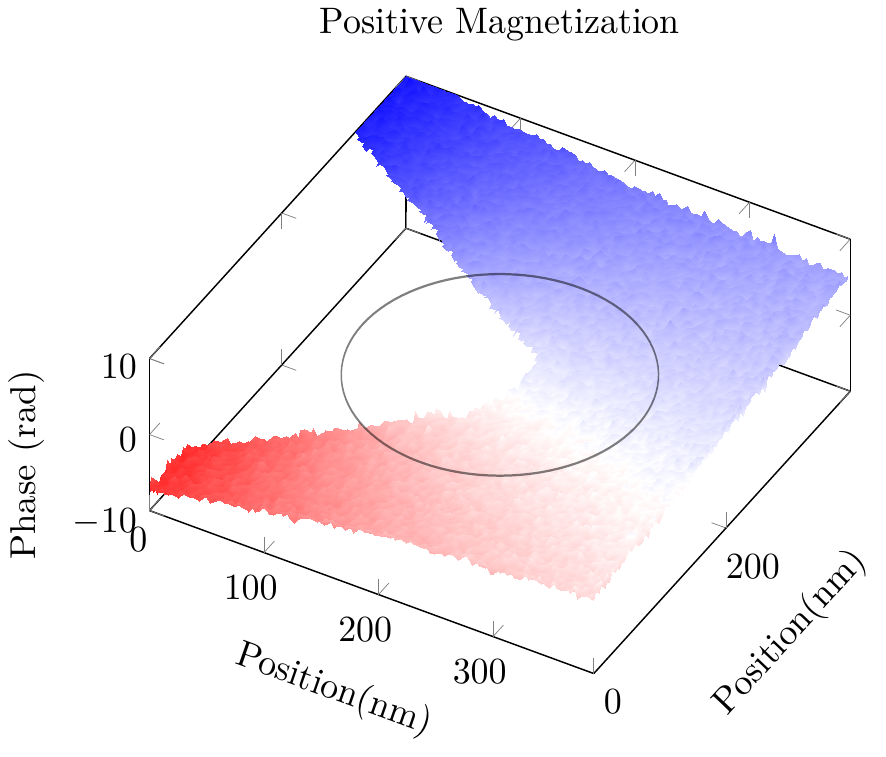}\label{fig:phase.ramp.exp}}\hfil
	\subfloat[]{\includegraphics[height=.2\textheight,clip=true,trim= 0 5 0 0]{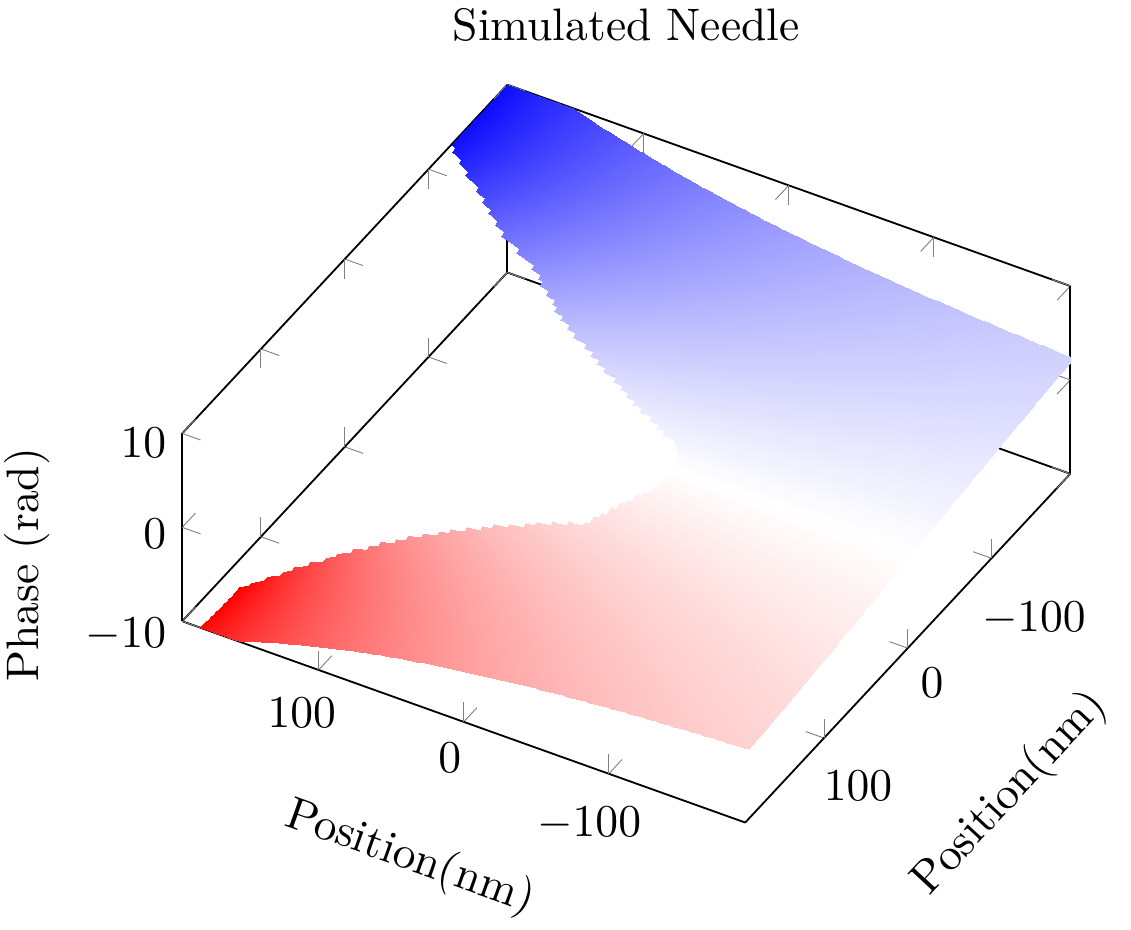}\label{fig:phase.ramp.simulation}}
	\subfloat[]{\includegraphics[height=.2\textheight]{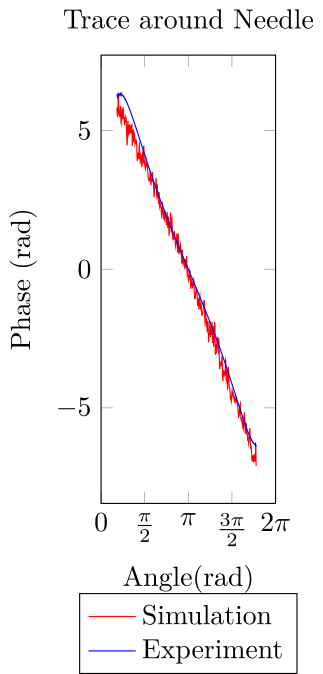}\label{fig:evaluation.of.charge}} 
	\caption{
	\textbf{Holographic measurement of the (unquantized) magnetic charge associated with 
	a magnetic needle.} 
	\textbf{(a)} Standard configuration of electron microscope used for holographic measurements; 
	an electron wave interacting with the specimen interferes with a reference beam 
	to create a hologram which encodes its phase (cf Fig. 2, Supplementary Information). 
	The sample used is shown in an inset.
	\textbf{(b)}  Ramp in phase measured in an electron wave traversing the tip of a magnetic needle.
	For greater clarity, the area masked by the needle has been excluded from the plot.
	\textbf{(c)} Ramp in phase found in equivalent finite-element simulation. 
	\textbf{(d)}  Phase on a closed contour encircling the tip of the needle 
	at a radius of $r=120\ \text{nm}$ (black circle in b), as found in simulation and experiment.
	The magnetic charge associated with the tip of the needle can be 
	estimated by integrating phase around this contour, following Eq.~\ref{eq:qm.spin.ice}.
	For a magnetic needle, the charge measured depends on the 
	radius of integration.
	For $r=120\ \text{nm}$, we find $q_m^{\rm exp} = (6.79 \pm 0.08) \times 10^{-9}\ \text{A.m}$, 
	in agreement with simulation.
	}
	\label{fig:holo}
\end{figure*}


One technique which has the potential to overcome this barrier is electron 
holography \cite{Tonomura1999}.
Early in the history of magnetic monopoles \cite{Curie1894}, it was 
realised that their interaction with electrons would have interesting 
properties \cite{Poincare1896}.
And with the arrival of quantum mechanics, it was 
established that an electron encountering a monopole 
acquires a characteristic discontinuity (branch cut) in the phase of its wave 
function \cite{Tamm1931,Fiertz1944}, with a jump in phase 
proportional to the magnetic charge of the monopole \cite{Wilczek1982}.
Holography, which uses interference to measure the phase of an 
electron wave \cite{Gabor1948}, therefore provides a way of both identifying 
individual monopoles, and measuring their magnetic charge.


Today, electron holograms are most commonly measured using a transmission 
electron microscope (TEM) equipped with an electrostatic biprism \cite{Tonomura1999} [Fig.~\ref{fig:microscope}].
Since the phase of electron waves is modified by the matter they pass 
through, 2D holograms encode information about 3D 
samples~\cite{Mankos1996,McCartney1997}. 
And where an electron wave interacts with a magnetic field, it 
picks up an additional phase through the Aharonov--Bohm 
effect \cite{Ahranov1961} 
\begin{eqnarray}
\phi = -\frac{e}{\hbar} \int A_z dz,
\label{eq:phiAB}
\end{eqnarray}
where $A_z$ is the magnetic vector potential in the direction of the 
electron's motion. 
This makes electron holography a powerful tool for imaging 
magnetic structures, from the macroscopic down to the atomic 
scale~\cite{Lichte2002,McCartney2005,Fert2013,Park2014,Zheng2017}.


{\bf Experiments on magnetic needles.}   
As a prelude to measurements on spin ice, we consider first 
a macroscopic analogue of a magnetic monopole, in the form of a magnetic needle,  
extending the earlier experiments of B\'ech\'e {\it et al}~\cite{Beche2014}.
We now show how these measurements can be used to determine magnetic charge.
In Fig.~\ref{fig:phase.ramp.exp}, we present the phase--map reconstructed from
holographic images of our sample [inset to Fig.~\ref{fig:microscope}].
This exhibits a characteristic ramp in phase, winding around the tip of the needle, 
and terminating in a line--like discontinuity, which corresponds to the Dirac string 
of a magnetic monopole~\cite{Beche2014}.  
Experimental results are found to be in quantitive agreement with the phase obtained from 
finite--element simulations of the field ${\bf A}({\bf r})$ generated by the magnetic 
sample~\cite{Beleggia2003} [Fig.~\ref{fig:phase.ramp.simulation}].
Moreover, as shown in Supplementary Information, this phase ramp vanishes when 
the magnetic needle is heated above its Curie temperature, confirming its magnetic origin.


The charge of a magnetic monopole can be found by integrating 
the phase of an electron wave function on a contour encircling the 
monopole \cite{Wilczek1982}
\begin{align}
	q_m = \frac{\hbar}{\mu_{0}e} \oint d\phi \; .
\label{eq:magnetic.charge}
\end{align}
We can apply this approach to a magnetic needle, with an important 
caveat: unlike a monopole, the magnetic needle is not a point--like object, 
and it does not have a quantized magnetic charge.  
It follows that the value of $q_m$ obtained will depend on the path of 
integration.
Choosing a circular path of radius $r=120\ \text{nm}$ [Fig.~\ref{fig:evaluation.of.charge}]
we find $q_m^{\rm exp} = (6.79 \pm 0.08) \times 10^{-9}\ \text{A.m}$, in 
good agreement with the value $q_m^{\rm sim} = 6.78 \times 10^{-9}\ \text{A.m}$ 
found in simulation.
	

\begin{figure*}
\centering
$\vcenter{
\hbox{
\subfloat[]{\includegraphics[width=.22\textwidth]{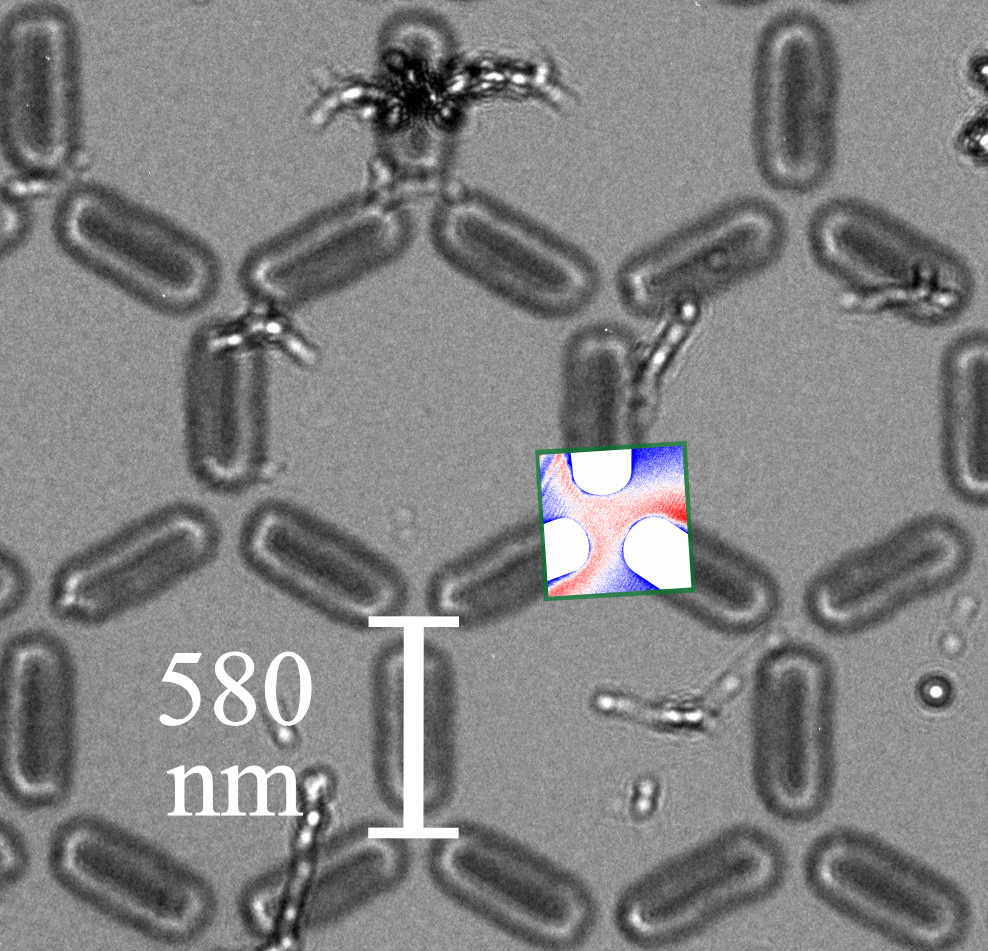}
\label{fig:artificial.spin.ice.sample}}
}}$
\begin{tabular}{|m{.37\textwidth}m{.37\textwidth}}
\subfloat[]{\includegraphics[height=.2\textheight]{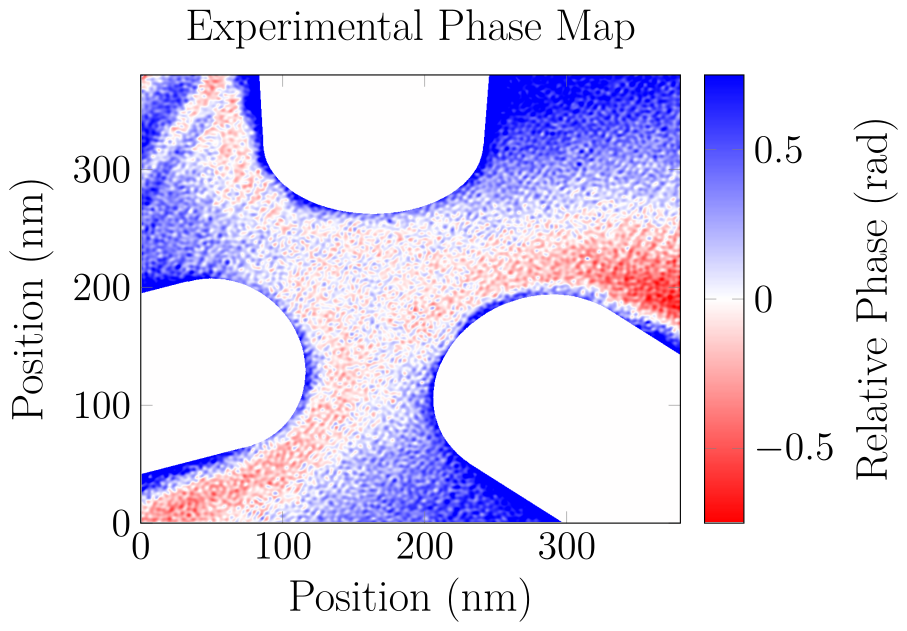}
\label{fig:phase.map.3q.vertex}} &
\addtocounter{subfigure}{1}\subfloat[]{\includegraphics[height=.2\textheight]{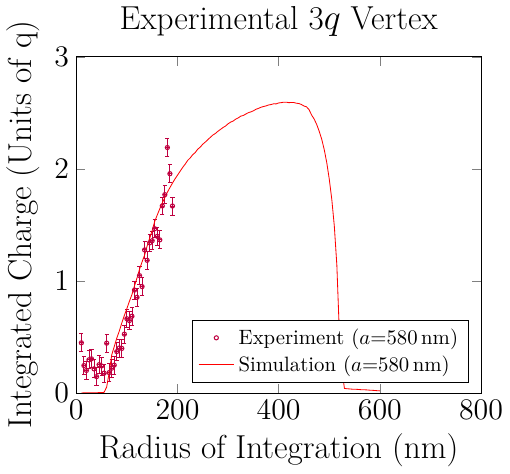}
\label{fig:integrated.charge.3q.vertex}}\\
\addtocounter{subfigure}{-2}\subfloat[]{\includegraphics[height=.2\textheight]{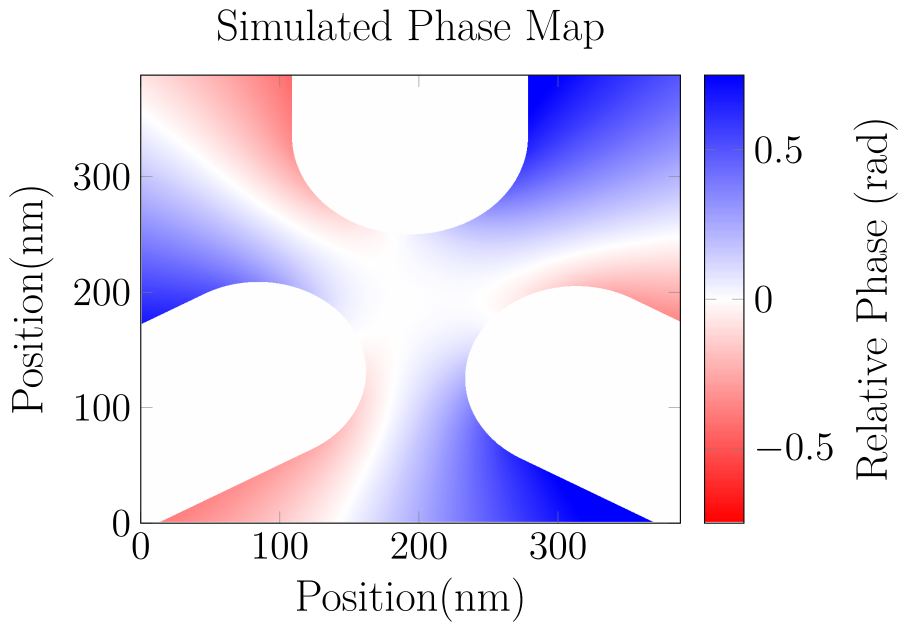}
\label{fig:simulation.3q.vertex}} & 
\addtocounter{subfigure}{1}\subfloat[]{\includegraphics[height=.2\textheight]{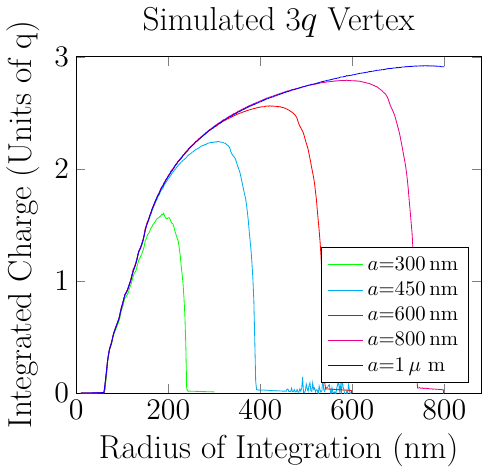}
\label{fig:integrated.charge.ideal}}\\
\end{tabular}
\caption{
	\textbf{Holographic measurements of quantized magnetic monopole with charge $3Q$ in artificial spin ice.} 
	\textbf{(a)} Defocused Lorentz--mode image of artificial ``Kagome'' spin ice, 
	comprising a honeycomb network of permalloy magnetic islands, with a lattice constant 
	of $580$ nm, on a SiO base.
	Coloured overlay shows the individual vertex resolved in (b).  
	\textbf{(b)} Phase map of a monopole excitation with a $+3q$ magnetic charge, reconstructed from TEM measurements.
	\textbf{(c)} Comparison to simulation for a $3q$--vertex.
	\textbf{(d)} Magnetic charge found by integrating phase around $3q$--vertex, measured 
	in units of $q_{m}^{\rm ASI}$, as a function of the radius of integration $r$. Experimental data (points with error bars) match quantitatively with simulations (solid line). The radius $r$ is limited by the field of view in TEM experiments.
	\textbf{(e)} Charge integrated for simulations of islands with varying lattice constant $a$, showing how the measured charge converges on $3q_{m}^{\rm ASI}$ in the limit of a long, thin island.
	For the sample studied, the natural unit of monopole charge is 
	$q_{m}^{\rm ASI} = 5\e{-10}\ \text{A$\cdot$m}$.
	}
\label{fig:artificial.spin.ice}
\end{figure*}


{\bf Experiments on artificial spin ice.}   
Having established our methodology, we now turn to the magnetic 
monopoles of an artificial spin ice system.
Artificial spin ices are constructed by fabricating a lattice of micron--scale 
magnetic islands, each of which acts as a mesoscopic magnetic moment, 
subject to the ice rules \cite{Wang2006,Moeller2006,Tanaka2006,Mengotti2008,Ladak2012,Mengotti2011}.
Dirac strings and monopoles in artificial spin ice have previously been studied using transmission electron microscopy, but with the limitation of utilizing numerical reconstruction of focal series stacks \cite{phatak11a,pollard12a,phatak18a}, or focus on quantification of island in-plane magnetization \cite{wessels22a}. 
Our motivation here is to go beyond the state of the art in this field and \textit{directly} visualise the phase shift due to emergent magnetic monopoles using off-axis electron holography, which will allow us a quantitative measurement of the magnetic charge, and offers the possibility of time-resolved observation of monopole dynamics. 
We work with ``Kagome spin ice''  \cite{Mengotti2008,Tanaka2006}
a honeycomb network [Fig.~\ref{fig:artificial.spin.ice.sample}], 
for which the ice rules are 2--in,1--out (1--in,2--out), such that 
vertices have (quantized) magnetic charge $\pm q$.
Meanwhile, monopole excitations have 3--in (3--out) spins 
with charge $\pm 3q$.
The sample we study has a lattice constant $a = 580\ \text{nm}$, and islands 
with magnetic moment $m = (2.9 \pm 0.3) \times 10^{-16} \text{A.m}^2$. 
Within a dumbbell model $m = qa$ \cite{Castelnovo2008}, this gives a 
natural scale of magnetic charge $q_{m}^{\rm ASI} = (5.0 \pm 0.4) \times 10^{-10} \text{A.m}$.


In Fig.~\ref{fig:artificial.spin.ice} we show the results of holographic 
measurements.   
We consider the phase map reconstructed from measurements of a monopole excitation with a $+3q$ charge (3--in) [Fig.~\ref{fig:phase.map.3q.vertex}], which compares remarkably well with simulations [Fig.~\ref{fig:simulation.3q.vertex}]. Resolving the phase on a circular path of radius $r$ around the vertex, we recognise three successive phase ramps -- from negative (red) to positive (blue) -- between each pair of islands. Integrating Eq.~(\ref{eq:magnetic.charge}) we obtain the magnetic charge $Q_{m}^{\rm exp}$ of a $3q$ monopole excitation on Kagome spin ice [Fig.~\ref{fig:integrated.charge.3q.vertex}].  For the largest radius $r \approx 200\ \text{nm}$ available in the experimental window of measurement, $Q_{m}^{\rm exp}$ reaches about $2/3$ of the theoretical expectation, $3q_{m}^{\rm ASI}= 15.0 \times 10^{-10} \text{A.m}$.


\begin{figure*}
	\centering
		\subfloat[]{\includegraphics[height=.2\textheight,clip=true,trim=0 0 80 0]{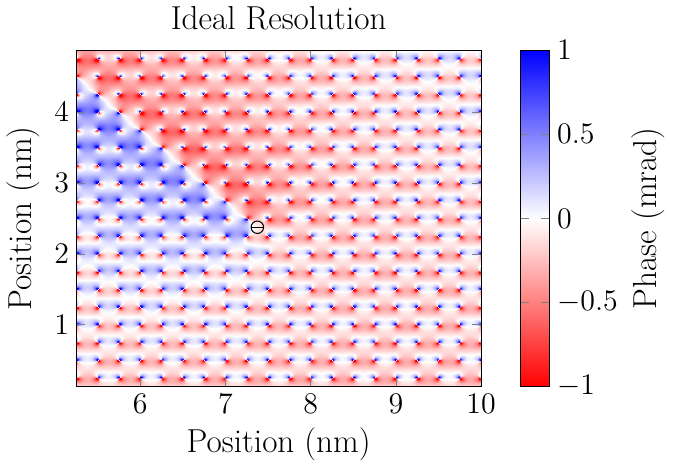}
			\label{fig:single.monopole.ideal.resolution}}
		\subfloat[]{\includegraphics[height=.2\textheight]{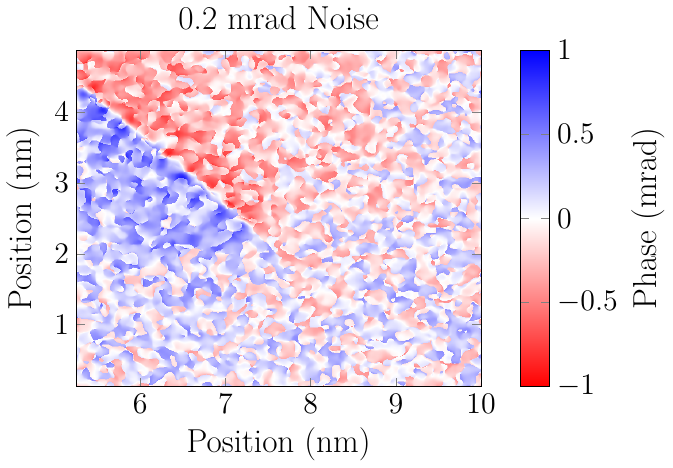}
			\label{fig:single.monopole.finite.resolution}}
		\subfloat[]{\includegraphics[height=.2\textheight]{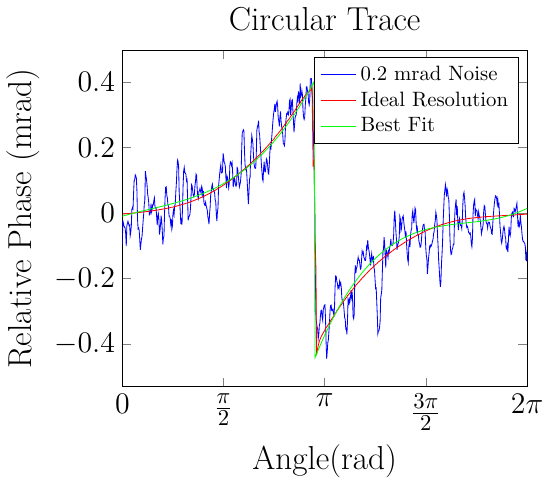}
			\label{fig:single.monopole.phase.ramp}}
	\caption{
	\textbf{Holographic signal of a single, quantized magnetic monopole in a thin film of spin ice, polarised by magnetic field.} 
	\textbf{(a)} Phase map found in simulations with a spatial resolution of $20\ \text{pm}$, 
	assuming perfect phase resolution.
	An isolated magnetic monopole ($\ominus$) can be identified at the end of a 
	Dirac string (diagonal white line), associated with a jump in phase of the 
	electron wave function, $\Delta \phi \sim 0.8\ \text{mrad}$.
	\textbf{(b)} 
	Equivalent phase map with a spatial resolution of $240\ \text{pm}$, 
	and phase noise $\pm0.1\ \text{mrad}$.
	The Dirac string, and associated phase jump, remain clearly visible.
	\textbf{(c)} 
	Phase ramp extracted from circular trace about magnetic monopole at a radius 
	of $2\ \text{nm}$.
	The magnetic charge of the monopole can be accurately determined from 
	the jump in phase across the Dirac string, even in the presence of noise.
	The monopole charge determined from fitting the blue data is $q_m = 4.4 \pm 0.3 \e{-13}\ \text{A.m}$, in very good agreement with the dumbbell model of spin ice \cite{Castelnovo2008} [Eq.~(\ref{eq:qm.spin.ice})] despite the presence of phase noise.
	Simulations were carried out on a film of spin ice, one unit-cell thick, 
	for parameters appropriate to Dy$_2$Ti$_2$O$_7$, with magnetic field 
	parallel to a cubic crystal axis in the plane of the image.  
	}
	\label{fig:single.monopole}
\end{figure*}


The resolution of this apparent mismatch lies in the use of the dumbbell model, 
which is only strictly valid for an infinitely long and thin magnetic island.
The role of finite aspect ratio is an aspect of the problem which can easily 
be studied in simulation. In Fig.~\ref{fig:integrated.charge.ideal} 
we show how estimates of charge $Q_{m}^{\rm sim}$ depend on radius of integration $r$ 
and lattice constant $a$, keeping other parameters constant.
For a given lattice of constant $a_{0}$, the initial increasing phase is the same for 
all lattices made of longer islands $a>a_{0}$; the charge does not ``see'' the end of 
the islands beyond the integration radius.
$Q_{m}^{\rm sim}$ passes a maximum before falling rapidly when $r$ reaches the end of the three islands.\footnote{Note that in our simulations, the islands start at 50 nm from the centre of the vertices.} 
The value of this maximum increases with $a$ and converge asymptotically to the saturated value $3q_{m}^{\rm ASI}$ as the islands become infinitely long. 
Note that the asymmetry of the curves -- smooth increase and sharp decrease -- is largely due to the presence of 12 surrounding islands in our simulations, whose inclusion is necessary to match simulations to experiments in Fig.~\ref{fig:simulation.3q.vertex}.


Our experiments on artificial spin ice confirm that it is possible to measure 
the charge of a monopole excitation using electron holography.   
And in contrast with the magnetic needle \cite{Beche2014}, this charge 
saturates at a quantized value. 
None the less, the ``point source'' of magnetic field is smeared over a region of size 
$a/2$, introducing finite--size effects which need to be taken into account when 
analysing data.
%

\begin{figure*}
	\centering
	\subfloat[]{\includegraphics[height=.165\textheight,clip=true,trim=0 0 90 0]{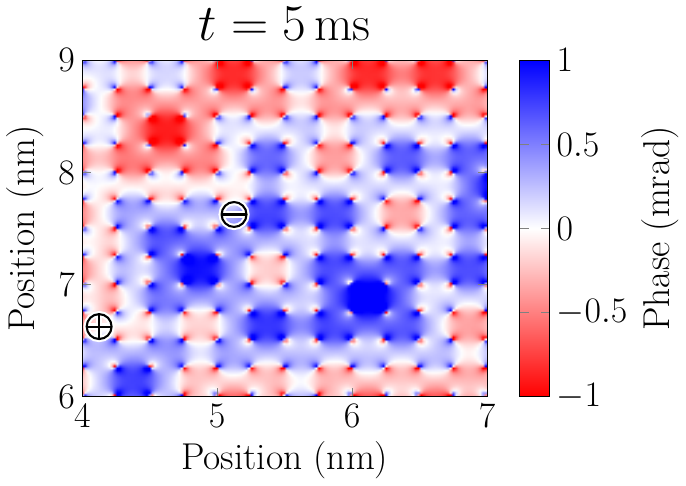}
		\label{fig:5a}}
	\subfloat[]{\includegraphics[height=.165\textheight,clip=true,trim=0 0 90 0]{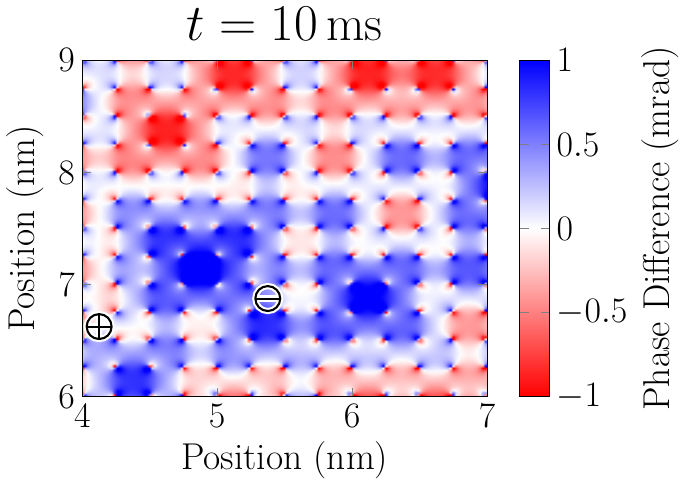}
		\label{fig:5b}}
	\subfloat[]{\includegraphics[height=.165\textheight,clip=true,trim=0 0 90 0]{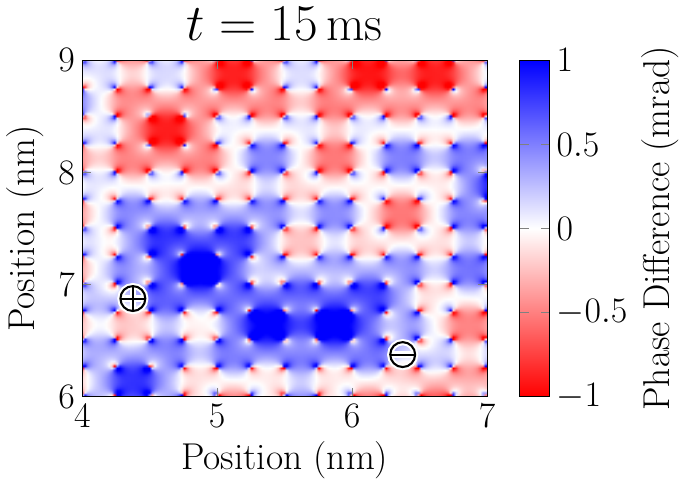}
		\label{fig:5c}}
	\subfloat[]{\includegraphics[height=.165\textheight]{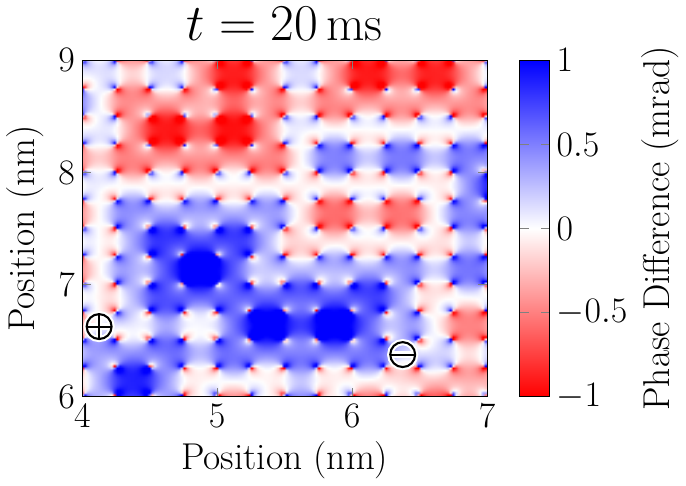}
		\label{fig:5d}}\\
	\subfloat[]{\includegraphics[height=.165\textheight,clip=true,trim=0 0 90 0]{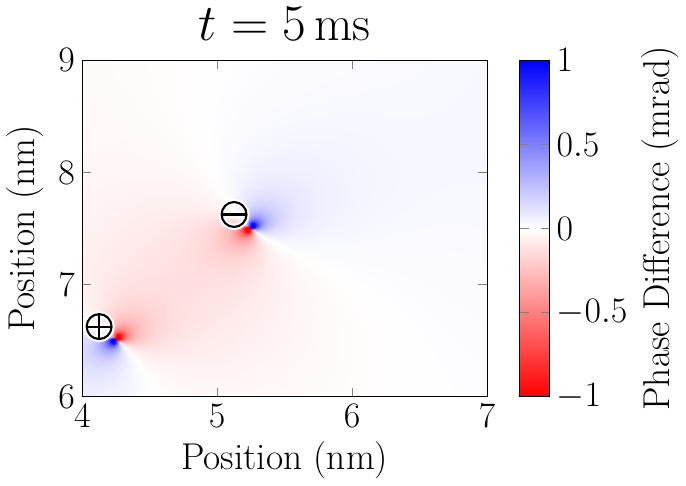}
		\label{fig:5e}}
	\subfloat[]{\includegraphics[height=.165\textheight,clip=true,trim=0 0 90 0]{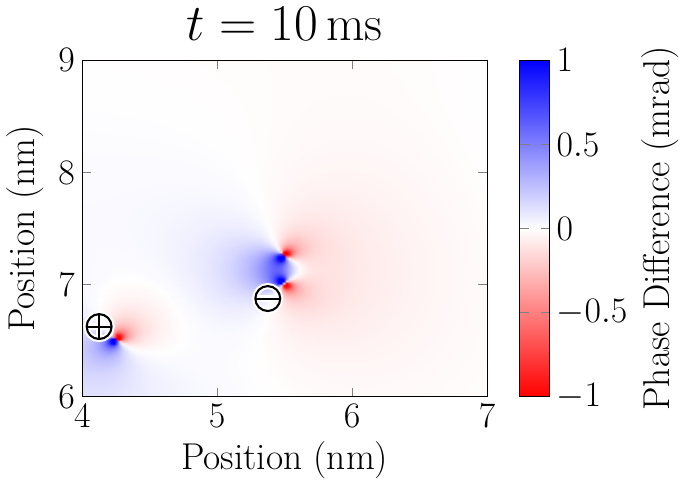}
		\label{fig:5f}}
	\subfloat[]{\includegraphics[height=.165\textheight,clip=true,trim=0 0 90 0]{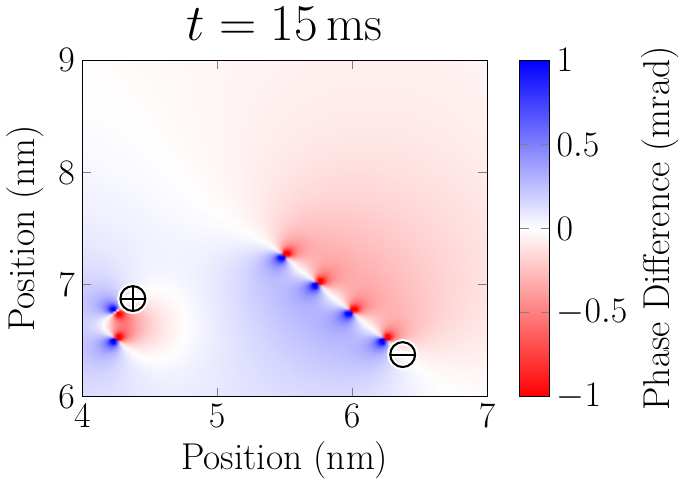}
		\label{fig:5g}}
	\subfloat[]{\includegraphics[height=.165\textheight]{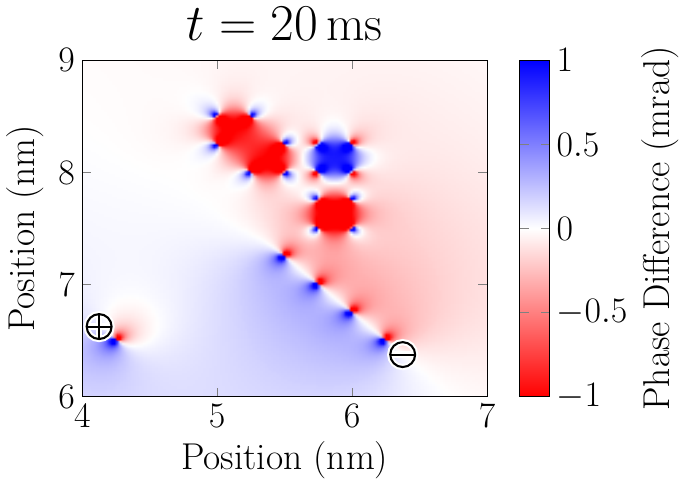}
		\label{fig:5h}}\\
	\subfloat[]{\includegraphics[height=.165\textheight,clip=true,trim=0 0 90 0]{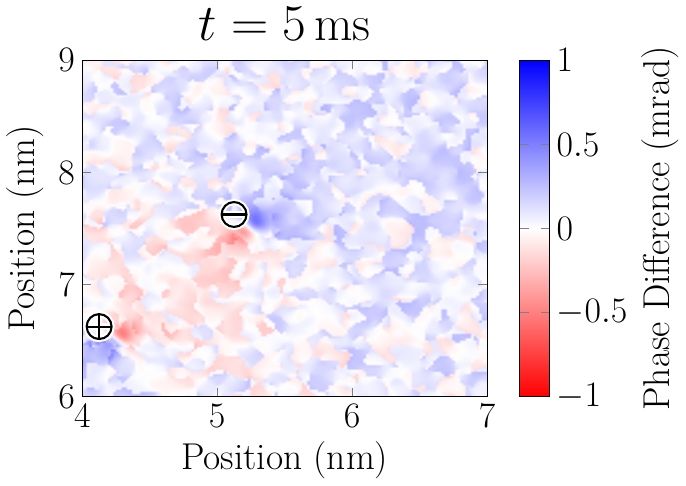}
		\label{fig:5i}}
	\subfloat[]{\includegraphics[height=.165\textheight,clip=true,trim=0 0 90 0]{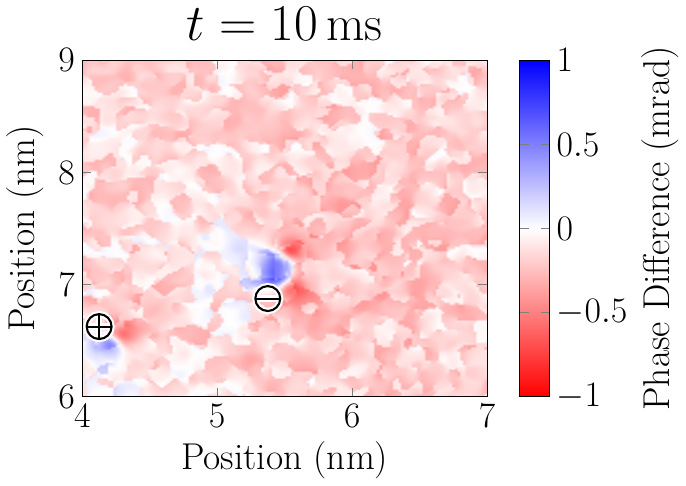}
		\label{fig:5j}}
	\subfloat[]{\includegraphics[height=.165\textheight,clip=true,trim=0 0 90 0]{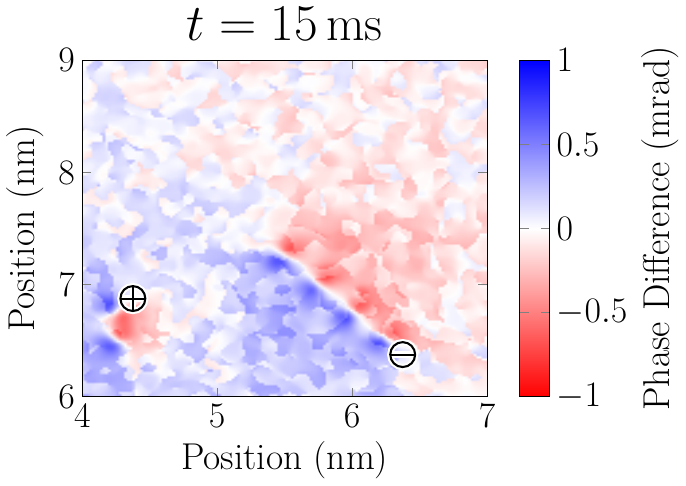}
		\label{fig:5k}}
	\subfloat[]{\includegraphics[height=.165\textheight]{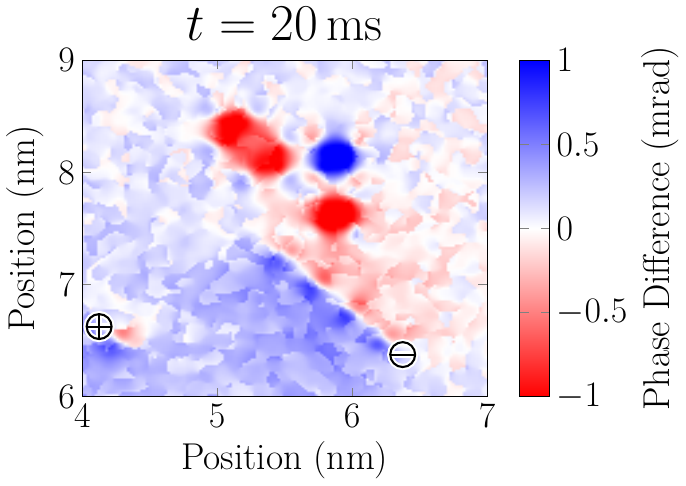}
		\label{fig:5l}}
	\caption{
	\textbf{Holographic signals of quantized magnetic monopoles, and emergent electric fields,  
	in a thin film of spin ice at \mbox{$T=700\ \text{mK}$}.} 
	\textbf{(a-d)} Phase maps taken from simulations at time stamps of 
	$t = 5, 10, 15, 20\ \text{ms}$, showing magnetic monopoles ($\ominus$/$\oplus$) subject to 
	diffusive dynamics.
	Results are shown for a spatial resolution of $20\ \text{pm}$, and ideal phase 
	resolution.
	\textbf{(e-h)} Differential phase maps for the spin configurations shown in (a-d), 
	calculated relative to phase at time $t =0$.
	Magnetic monopoles trace Dirac strings as they move.  
	Where monopoles annihilate after traversing a closed loop, the associated fluctuation 
	of an emergent electric field is visible as a region of constant phase difference.  
	\textbf{(i-l)} Differential phase maps equivalent to (e-h), but calculated for a spatial 
	resolution of $240\ \text{pm}$, and with phase noise of $0.1\ \text{mrad}$.
	Fluctuations of electric field remain clearly visible.
	Movement of magnetic monopoles can be distinguished by comparing successive 
	frames in simulations [cf.~Animation in Supplementary Information].
	Simulations were carried out for parameters appropriate to Dy$_2$Ti$_2$O$_7$, 
	with a cubic lattice constant of $a = 1.00(1)\ \text{nm}$, for a film $1\ \text{nm}$ thick, 
	in the absence of magnetic field. 
	}
	\label{fig:spin.ice.700mK}
\end{figure*}


{\bf Simulation of pyrochlore spin ice.}  
In pyrochlore spin ice, 
monopoles are localised on a scale of a single tetrahedron, $\sim 3.5\ \text{\AA}$.
This is a favourable scale for observation using an electron with wavelength  
$\lambda \sim 0.2\ \text{\AA}$, typical of a modern TEM.
Moreover, the motion of monopoles in spin ice occurs on timescales 
$\sim 1\ \text{ms}$ \cite{Jaubert2009}, comparable with the 
rate at which holographic images can be taken.
This raises the prospect of directly imaging a magnetic 
monopole, and its associated dynamics, by the methods 
applied to artificial spin ice.
Since available TEMs are not equipped to operate at temperatures 
$\lesssim 1\ \text{K}$, relevant to spin ice, we have used simulation 
to explore what could be learned from holographic experiments 
on a thin film of pyrochlore spin ice.
For concreteness, we consider films with the thickness of 
a single (cubic) unit cell, and parameters appropriate to Dy$_2$Ti$_2$O$_7$, 
which has a (cubic) lattice constant 
$a = 10.0 (1)\ \text{\AA}$ \cite{denHertog2000}.
%


We start by considering the limit which is easiest to understand; 
a system polarised by magnetic field \cite{Morris2009}, 
and containing a single monopole [Fig.~\ref{fig:single.monopole}].
In this case the excitation comprises a chain of flipped spins (Dirac string), 
terminating in a tetrahedron with a 3--in, 1--out configuration 
(magnetic monopole), within a state where all other tetrahedra obey 
the ice rules ($\nabla \cdot {\bf H} = 0$), and are polarised in the 
plane of the sample.
These features are immediately obvious in the phase map 
[Fig.~\ref{fig:single.monopole.ideal.resolution}], where an isolated 
magnetic monopole ($\ominus$) can be identified at the end of a 
Dirac string (diagonal white line), associated with a jump in phase 
of the electron wave function, $\Delta \phi \sim 0.8\ \text{mrad}$.
Being topological in character, these features are 
stable against the effects of finite spatial and phase 
resolution [Fig.~\ref{fig:single.monopole.finite.resolution}].
And estimating the charge of the magnetic monopole is  
now straightforward; integrating a circle 
of radius $r = 2\ \text{nm}$, 
in the presence of phase noise $0.2\ \text{mrad}$ 
[Fig.~\ref{fig:single.monopole.phase.ramp}], 
we find $q_m  = (4.4 \pm 0.3)\times 10^{-13}\ \text{A.m}$, 
in good agreement with the expected value [Eq.~\ref{eq:qm.spin.ice}].


We now turn to a thermalised sample, in the absence of magnetic field.
In Fig.~\ref{fig:5a}--\ref{fig:5d} we show simulated phase maps for spin 
configurations drawn 
from Monte Carlo (MC) simulation 
of a sample at $T = 700\ \text{mK}$.
In regions without monopoles, e.g. top right corner of Fig.~\ref{fig:5a}, 
the phase map can be divided into discrete ``boxes'' with phase 
$\phi \sim \pm 0.4\ \text{mrad}$, a fact related to the 
``height representation'' of two--dimensional vertex models \cite{Baxter1989}.   
However in regions visited by monopoles (remainder of sample), phase 
maps are more complicated, reflecting the fact that the Dirac string connecting 
monopoles cannot be uniquely defined in the absence of magnetic field \cite{Morris2009}.


The key to imaging monopoles in thermalised samples  
is to consider the difference in phase which accumulates over time, 
relative to an initial spin configuration.
Since all changes in spin configurations come from the movement of 
monopoles, this provides a direct measurement of both the location 
of monopoles, and the path by which they have moved.   
Moreover, it is known that the Markovian dynamics 
of classical MC  simulation, carried out using local spin updates, gives a good 
account of the diffusive dynamics of monopoles in spin ice \cite{Jaubert2009,Jaubert2011-JPCM23}.
This implies that simulation time can be used as a proxy for real time, 
up to a known conversion factor of $1\ \text{MC step} = 2.5\ \text{ms}$, 
characteristic of Dy$_2$Ti$_2$O$_7$ at $T = 700\ \text{mK}$ \cite{Jaubert2011-JPCM23}.


In Fig.~\ref{fig:5e} we show a map of the phase difference at a timestamp 
of $t = 5\ \text{ms}$, relative to a reference state 
at $t = 0$, 
in which two monopoles were present.
(See Supplementary Information for an 
animation of the phase--difference as a function of time).  
During the first $5\ \text{ms}$, 
both monopoles have moved a short distance, leaving  
a track which functions much like the Dirac string for a monopole in field, 
with a ramp in phase $\Delta \phi \sim 0.8\ \text{mrad}$ marking the 
(change in) magnetic charge at each end of the track.
Qualitatively similar results are found at $t = 10\ \text{ms}$ [Fig.~\ref{fig:5f}]
and $t = 15\ \text{ms}$ [Fig.~\ref{fig:5g}], as the two monopoles continue 
to diffuse around the lattice.
And being topological in nature, the tracks left by moving monopoles 
remain discernable  in the presence of finite phase noise 
[Fig.~\ref{fig:5i}, Fig.~\ref{fig:5j}, Fig.~\ref{fig:5k}].


A qualitatively new feature is visible in the phase--difference maps 
for $t = 20\ \text{ms}$ [Fig.~\ref{fig:5h}, Fig.~\ref{fig:5l}].
In addition to tracks left by moving monopoles, these exhibit bounded regions 
of constant phase--difference $\Delta\phi \sim \pm 0.4\ \text{mrad}$, with both 
square and hexagonal perimeters.
These reflect (incoherent) tunnelling between different states obeying the ice rules.
Over a timescale $\delta t \sim 5\ \text{ms}$ ($2\ \text{MC steps}$) it is possible 
for a pair of monopoles to come into existence, traverse a closed loop of spins, 
and anihilate. 
The shortest path on which they can do so is one of hexagonal plaquettes of the 
pyrochlore lattice [cf. Fig.~\ref{fig:electric.field} and the hexagonal pattern of Fig.~\ref{fig:5h}]. The square patterns in Fig.~\ref{fig:5h} are recognised as two-dimensional projections of a four-spin spiral connecting the top to the bottom surfaces of the thin film.


Associated with these new features, is a new piece of physics.
Where an electric current is transported around a loop, it acts as a source of 
magnetic field (current loop).   
By direct analogy, where a magnetic charge travels around a loop, this has 
the interpretation of a magnetic current loop, i.e. a source of electric field.
Fluctuations of electric field have already been widely discussed in the 
context of quantum spin ice, where they are associated with coherent quantum 
tunnelling, and contribute  to emergent photon excitation~\cite{Hermele2004,Benton2012}.
To the best of our knowledge, the incoherent fluctuations of electric fields 
in a classical spin ice with diffusive monopole dynamics have not
previously been discussed.
None the less, they are a robust feature of our simulations of thin films 
of Dy$_2$Ti$_2$O$_7$.
And, since they involve a jump in phase over an extended 
area, electric fields should prove easier to observe in 
experiment than point--like magnetic monopoles, and 
their Dirac strings.
%


{\bf Conclusion.}
We have used electron holography to characterise the magnetic monopoles 
of spin ice, through experiments on artificial, ``Kagome spin ice'', and simulations 
of thin films of pyrochlore spin ice. 
Our experiments on artificial spin ice demonstrate how holography can be used 
to observe a magnetic monopole, measure its (quantized) magnetic charge, 
and estimate its effective size. 
They also offer new insights into the distributions of magnetic fields on the scale of the lattice, and the limits of the ``dumbbell model'', when applied to 
macroscopic magnetic moments.


Meanwhile, our simulations of pyrochlore spin ice provide a clear road map 
for observing individual magnetic monopoles and their dynamics, using 
electron holography. 
We exhibit phase maps for individual monopoles in a sample polarised 
by magnetic field, demonstrate how these can be used to measure 
magnetic charge, and show how time--resolved experiments could 
be used to isolate the signal of monopoles in a thermalised sample 
of spin ice. 
These results also reveal that electron holography is sensitive to the 
emergent electric fields of a spin ice with dynamics.


Having characterised pyrochlore spin ice through simulation, 
we conclude with a few remarks about what it would take to observe 
monopoles in experiment.
The synthesis of thin films of spin ice is now well established.~\cite{Bovo2014,Leusink2014}
Installing a sample stage in a TEM capable of holding a specimen at 
$T = 700\ \text{mK}$, while novel, presents no problem of principle.
The required spatial resolution ($\approx 350\ \text{pm}$ 
for a monopole; $\approx 700\ \text{pm}$ for a fluctuation of electric field) 
compares with a current state of the art of $240\ \text{pm}$.~\cite{Tanigaki2019}
And the temporal resolution $\gtrsim 5\ \text{ms}$ needed to see 
fluctuations of electric field is also easily achievable with modern 
electronics.
The greatest challenge, therefore, is likely to be phase resolution, 
with key phenomena occurring on a scale $\Delta \phi \sim 0.4\text{--}0.8\ \text{mrad}$.
Even so, this requires only an incremental improvement on a  
current benchmark figure of $1\ \text{mrad}$.~\cite{Suzuki2012,Tanigaki2019} 
We conclude that, with modest development of instrumentation, 
electron holography holds the realistic promise of directly imaging 
magnetic monopoles in pyrochlore spin ice, and their dynamics.\\

\textbf{Acknowledgements}
The authors acknowledge helpful conversations with Robert Baughmann, 
Steve Bramwell, Martha McCartney, Hyun Soon Park and Sameer Wagh. 
They also acknowledge support for sample fabrication from Hiroyuki Kuwae 
and Jun Mizuno.
This work was supported by the Theory of Quantum Matter Unit and the 
Quantum Wave Microscopy Unit of the Okinawa Institute of Science 
and Technology Graduate University (OIST). 
L.D.C.J. acknowledges financial support from CNRS (PICS No. 228338) and the ``Agence Nationale de la Recherche'' under Grant No.ANR-18-CE30-0011-01.\\

\textbf{Author contributions}\\
A.D.\ carried out all experiment and simulations, and contributed to the 
interpretation of experimental and simulation results.
L.D.C.J.\ co--supervised theoretical aspects of the project, wrote simulation 
codes for thin films of spin ice, and contributed to the interpretation of 
experimental and simulation results.
C.C.\ advised and assisted with experimental aspects of the work.
T.S.\ supervised experimental aspects of the project.
N.S.\ suggested the project, co--supervised all theoretical aspects,  
and contributed to the interpretation of experimental and simulation results.
All authors contributed to the writing of the manuscript.\\

\textbf{Competing interests}
The authors declare no competing interests.\\

{\Large{\bf Methods}}

\textbf{Holographic} measurements and characterisation were performed using a ThermoFisher Scientific Titan G2 300kV Transmission Electron Microscope (TEM) equipped with an electrostatic biprism~\cite{Cassidy2017}. The magnetic samples were mounted (needles) or patterned (artificial spin ice) onto standard 3mm TEM grids, before loading into standard side-entry TEM holders for insertion into the microscope. Images and holograms were acquired in Lorentz mode, in order to keep the region around the sample free of external magnetic fields. The biprism was mounted into the selected area aperture strip, and this could be loaded into view after alignment and focusing was complete. Holograms were reconstructed into phase maps through the use of Holoworks 5 as part of the Gatan Digital Micrograph software suite.\\

\textbf{Artificial Spin Ice} (ASI) was thermalised ex-situ based on the standard protocol of Ref.~\onlinecite{Wang2007}. Then, to induce the formation of all-in/all-out vertices after loading into the microscope, an external field was applied to the sample in situ~\cite{Ladak2010,Mengotti2011}. This was done by tilting the sample 30$^\circ$, applying the objective lens field to saturate the artificial spin ice along one direction (which requires about 5\% of the objective lens field, corresponding to 0.8kG flux density), and then applying a slightly weaker (3\%) lens field in the opposite direction to partially reverse this saturation. This forms $3q$ monopoles along the boundary between oppositely magnetized regions, allowing them to be reproducibly generated. 

The lattice spacing $a$ was measured through focused Lorentz imaging, $a=580\pm11\  \text{nm}$. Additionally, in order to screen the artificial spin ice lattice for suitable vertices to image, the general orientation of each magnetic island domain was inferred through \textit{defocused} Lorentz imaging [Fig.~3a].  As the electron beam experiences a Lorentz deflection in a direction perpendicular to the direction of magnetization, the presence of bright and dark fringes along the long edges of the magnetic islands allows the magnitude and direction of magnetization to be inferred~\cite{Mankos1996}.  \\

{\bf Monte Carlo simulations of a thin film of spin ice.} 
The model system chosen for pyrochlore spin ice simulations is dipolar spin ice with nearest neighbour exchange and dipolar interactions, using parameters for Dy$_2$Ti$_2$O$_7$~\cite{denHertog2000}. The system is a slab of pyrochlore grown along the [001] axis, of one cubic unit-cell thick ($\approx 1.0$ nm); it includes three layers of tetrahedra and four layers of spins. All nearest-neighbour exchange couplings are included, including the orphan bonds on the surfaces that do not belong to tetrahedra \cite{Jaubert2017}. Such a system could be realised in heterostructures where the spin-ice thin film is sandwiched between two non-magnetic pyrochlore lattices. All spins interact with each other via magnetic dipolar interactions. The Ewald summation for dipolar interactions has been adapted to the slab geometry \cite{Jaubert2017}.

The in-plane system size is $L=10$ cubic unit cells, which means the total number of spins is $N=16\, L^{2}=1600$. Monte Carlo simulations are first thermalised at $T=700$ mK, where monopoles are present but sparse~\cite{Jaubert2011-JPCM23}. During thermalisation \cite{Jaubert2017}, we use a combination of worm algorithm  \cite{Melko2004} and Metropolis single-spin-flip updates. After thermalisation, measurements are taken using Metropolis single-spin-flip updates, which are known to approximate the dynamics of spin-ice compounds \cite{Jaubert2009}. One Monte Carlo step (MCs) is $N$ attempts to flip a spin, and corresponds to $2.5$ ms in real time~\cite{Jaubert2011-JPCM23}. \\

{\bf Computation of the vector potential} was accomplished through the use of a parallelised Fortran code to sum the vector potential in a 3D volume around a sample, with spatial resolution of 10\,pm in each dimension. In the case of the needle and artificial spin ice, the magnetic objects (needle and islands) were broken into small domains with uniform magnetic moment $m_{j}$.  From there the total vector potential in free space is
\begin{align}
\textbf{A}(\textbf{r}_i)&=\frac{\mu_0}{4\pi}\sum_{j}\frac{\textbf{m}_j\times \textbf{r}_{i,j}}{|\textbf{r}_{i,j}|^3}.
\label{eq:vecpot}
\end{align}

For pyrochlore spin ice, $\textbf{m}_{j}$ corresponds to Ising spins obtained from Monte Carlo simulations. To minimise boundary effects, periodic boundary conditions are used and Eq.~\ref{eq:vecpot} becomes
\begin{align}
\textbf{A}(\textbf{r}_i)&=\frac{\mu_0}{4\pi}\sum_{j,n}\frac{\textbf{m}_j\times \left(\textbf{r}_{i,j}+L(n_{x}\hat{\textbf{x}}+n_{y}\hat{\textbf{y}})\right)}{|\textbf{r}_{i,j}+L(n_{x}\hat{\textbf{x}}+n_{y}\hat{\textbf{y}})|^3},
\label{period}
\end{align}
where $L=10$ nm is the lattice period. We chose the number of periods in order to keep the relative error between $n$ and $n+1$ less than 0.1\%. We found $-5 \leq n_{x},n_{y} \leq 5$.\\

{\bf Spatial Resolution and Phase Noise Estimates}
In order to properly estimate the effects of lower spatial resolution and phase noise, spatial filters are applied to the high resolution phase maps. The reduced spatial resolution is generated by limiting the spatial information that can be reconstructed through holography. This is done by Fourier transforming the phase map, selecting only the information that represents length scales above the resolution limit via a circular aperture formed from a Gaussian window, and inverse transforming back into real space, represented as
\begin{align}
	I_{filtered} = \mathcal{F}^{-1}\left[\mathcal{F}[e^{-i\phi_{ideal}}]\left(|\textbf{k}-\textbf{k}_{edge}|\leq \frac{2\pi}{\textbf{x}_{res}}\right)\right],\label{eq:filter}
\end{align}
where $\textbf{x}_{res}$ is the spatial resolution limit, and $\textbf{k}_{edge}$ represents the FWHM of the Gaussian window. The resultant image is then taken as the filtered phase map. The phase noise is generated as shot noise from a Poisson distribution and then run through the same spatial filtering before being added to the filtered phase map. The standard deviation of this resultant noisy phase map dictates the phase resolution, as with experimental phase maps.\\

{\Large \bf References}
\bibliographystyle{apsrev4-1}
\bibliography{Manuscript}

\end{document}


\title{Supplementary Information}
	\author{Ankur Dhar}
	\email{ankur.dhar@alumni.oist.jp}
	\affiliation{Quantum Wave Microscopy Unit, Okinawa Institute of Science and Technology}
	\affiliation{Theory of Quantum Matter Unit, Okinawa Institute of Science and Technology}
	\author{L.\ D.\ C.\ Jaubert}
	\email{ludovic.jaubert@cnrs.fr}
	\affiliation{Theory of Quantum Matter Unit, Okinawa Institute of Science and Technology}
	\affiliation{CNRS, Universit\'e de Bordeaux, LOMA, UMR 5798, 33400 Talence, France}
	\author{Cathal Cassidy}
	\email{c.cassidy@oist.jp}
	\affiliation{Quantum Wave Microscopy Unit, Okinawa Institute of Science and Technology}
	\author{Tsumoru Shintake}
	\email{shintake@oist.jp}
	\affiliation{Quantum Wave Microscopy Unit, Okinawa Institute of Science and Technology}
	\author{Nic Shannon}
	\email{nic.shannon@oist.jp}
	\affiliation{Theory of Quantum Matter Unit, Okinawa Institute of Science and Technology}

	\date{\today}
	
	\maketitle
	\section{Holography Methods}
\begin{figure}[h]
		\centering
		\subfloat[]{\includegraphics[height=.35\textheight]{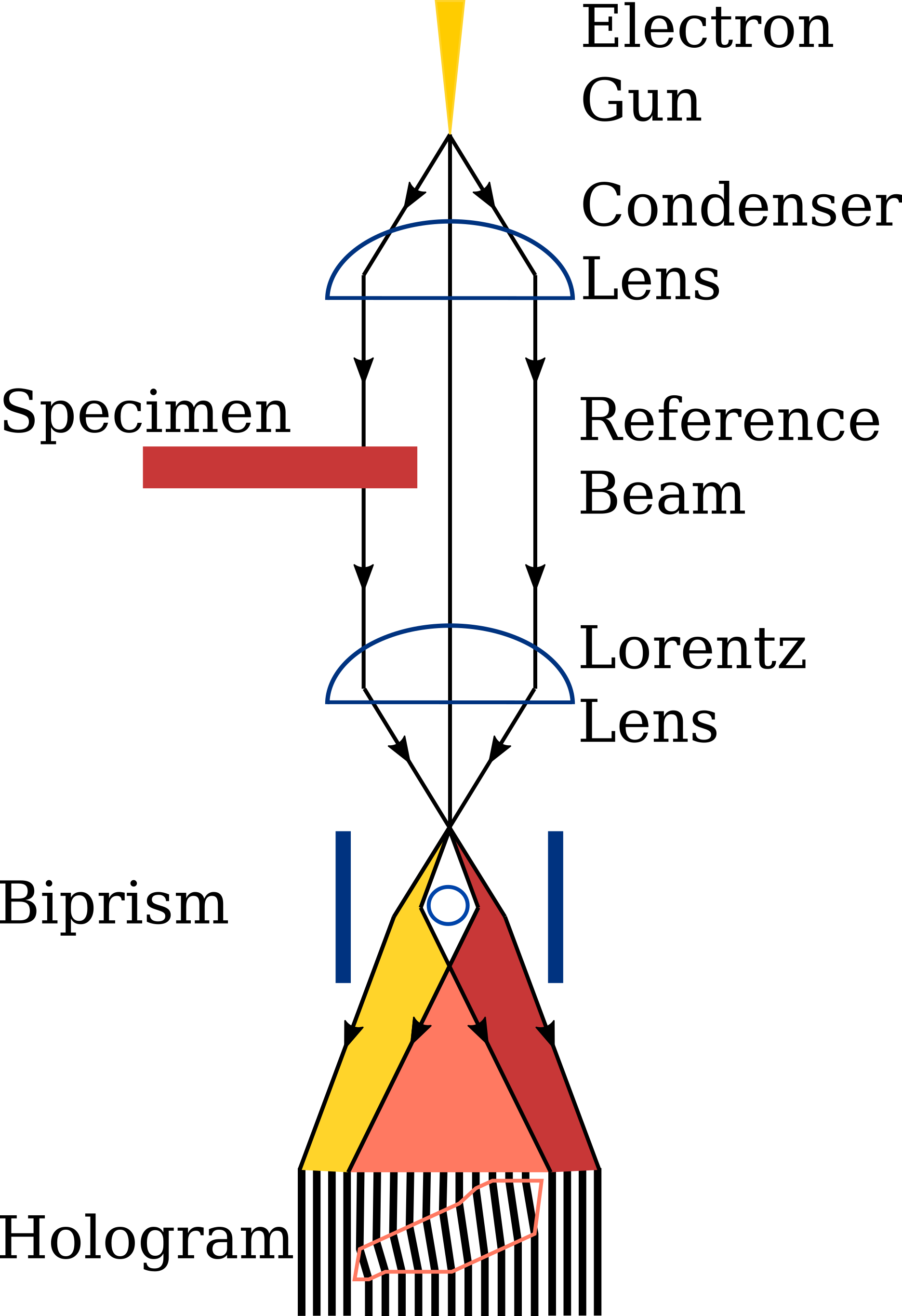}}
		\subfloat[]{\includegraphics[height=.35\textheight]{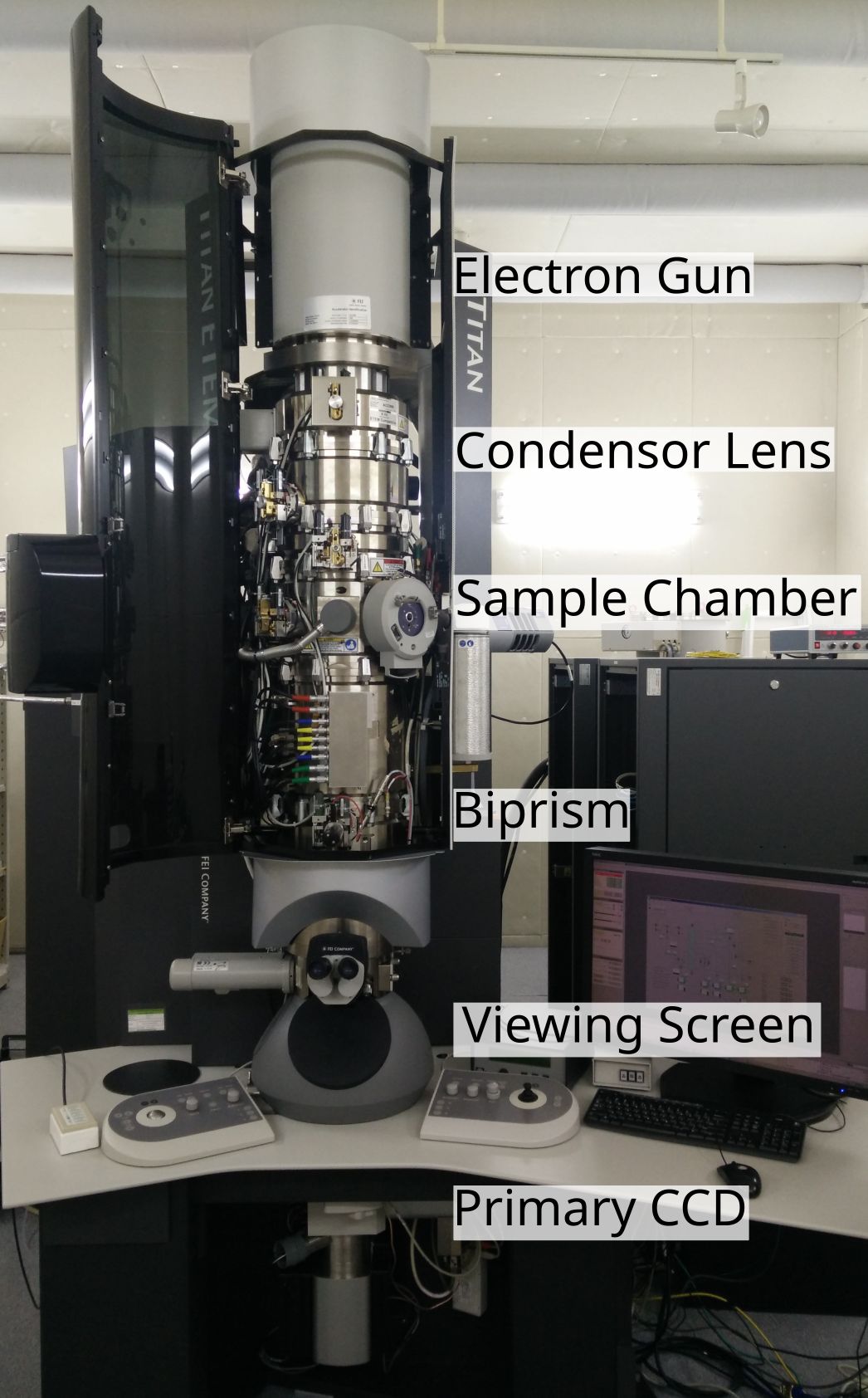}}
		\caption{\textbf{Overview of the main elements used for off-axis electron holography} \textbf{(a)} The elements of transmission electron microscope utilized for holography, including a Lorentz lens which is used in place of an objective lens for magnetic samples.  \textbf{(b)} Similar elements represented in the Titan microscope used for this study.}
		\label{fig:ETEM}
	\end{figure}

In Figures 2 and 3 of the main text, we present experimental electron holography measurements, acquired from magnetic needles and artificial spin ice, as magnetic monopole analogues. These measurements were carried out using a Thermofisher Scientific Titan G2 ETEM, operated at 300kV, as shown in Figure~\ref{fig:ETEM}.  This microscope was equipped with a Schottky (XFEG) field emission electron gun, S-TWIN pole piece with 5.4mm gap, Lorentz lens (for field-free imaging), post-specimen Cs-corrector (CEOS Gmbh) for the imaging forming lens system, and 2k x 2k Gatan Ultrascan XP1000 CCD camera.  To enable off-axis hologram formation, an electrostatic biprism was mounted in the Selected Area plane.  This experimental configuration allowed hologram fields of view of approximately 400nm x 400nm, and typical biprism fringe contrast of 20\% in vacuum.  The specified spatial resolution in Lorentz mode was 1.3nm.  For specimen mounting, a Thermofisher Scientific double-tilt low-background holder, and a Gatan 626 heating holder, were utilized, both of which accepted standard 3mm diameter TEM grids.  To study magnetic samples, the microscope was set into Lorentz mode before loading the sample, to ensure field-free conditions around the sample at all times.   Using Lorentz mode allows the electron wave emitted by the field emission gun in the microscope to propagate though the sample under field-free conditions, ensuring  that any change in phase is caused by the sample, as shown in Figure~\ref{fig:ETEM}. This phase difference between the sample beam and the reference beam (which remains unperturbed and propagates off-axis from the sample beam), is then encoded into an interference pattern, creating the measured hologram. 

\begin{figure}[h]
		\includegraphics[width=0.8\textwidth]{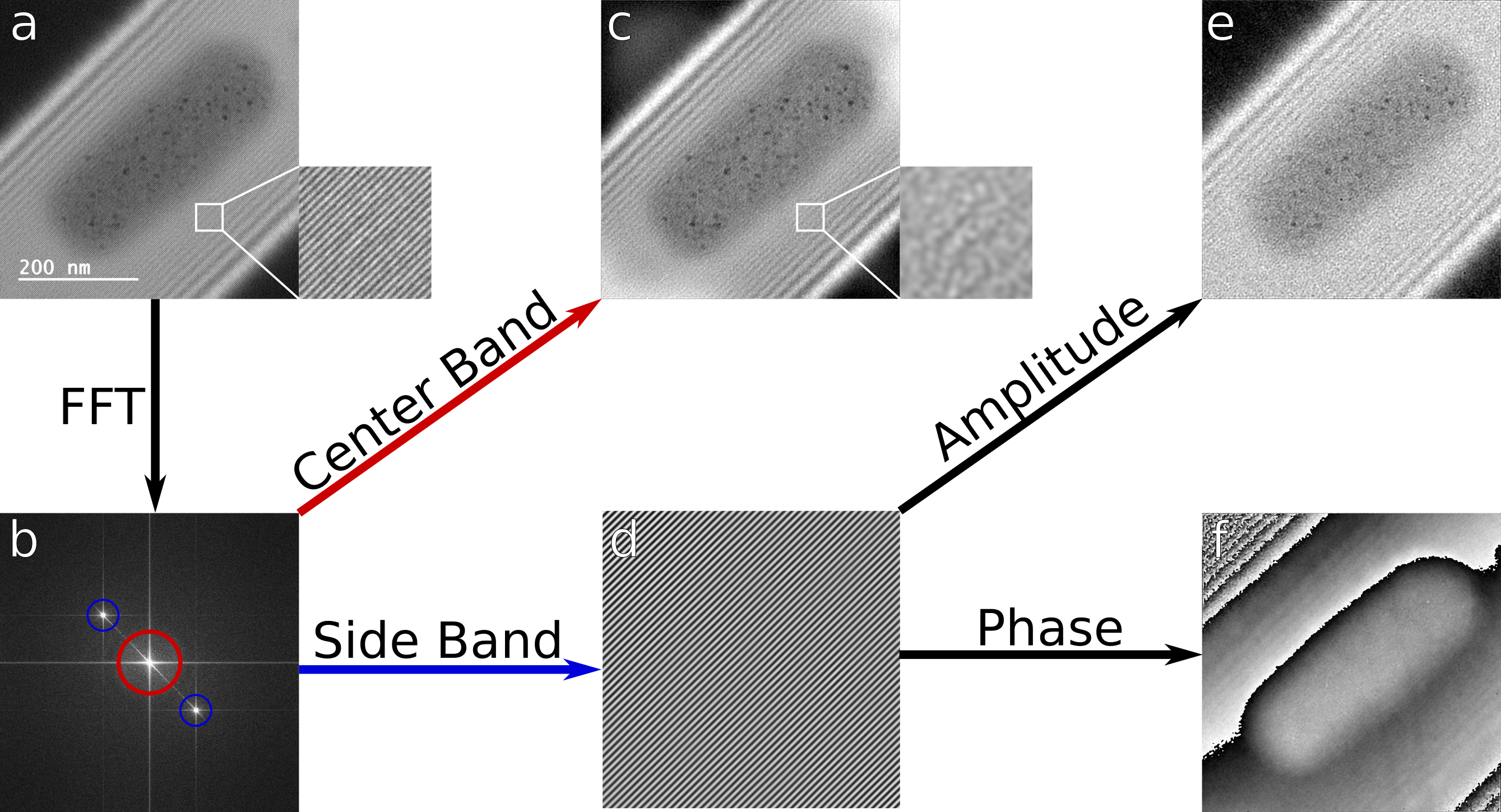}
		\caption{\textbf{Breakdown of the information extracted from a hologram.} Consider a hologram \textbf{(a)} of a magnetic island in an artificial spin ice sample. The primary interference fringes of the hologram, which encode the amplitude and phase information of interest, are present in the central region of the bright band (shown in zoomed inset). At the edges of the bright band, variable period Fresnel fringes arising from diffraction at the edges of the biprism wire, are evident.  These are an unavoidable artifact of the utilized experimental system. The corresponding Fourier Transform is shown in \textbf{(b)}. The sidebands (blue) arise from the primary fringes in the hologram, and constitute the primary carrier frequency for the amplitude and phase information. The centreband (red) arises from the conventional image intensity.  Note that the oblique streak through the origin arises from the Fresnel fringes, as well as the incomplete occupancy of the bright interference region in the acquired field of view - a large spread of Fourier frequencies is necessary to reproduce the sharp bright/dark transition at the edge of the bright band. Similarly, the horizontal and vertical streaks arise from the abrupt edges of the image.  For reconstruction of the data, selecting the centreband and performing an inverse Fourier Transform returns a real intensity image, similar to a normal TEM image \textbf{(c)}.  However, a selective  inverse Fourier Transform on the sidebands instead returns the actual interference fringes (zoomed) encoding the complex attributes of the electron wave (\textbf{d}). For standard hologram reconstruction, selection and translation of a single sideband to the origin in Fourier space, prior to performing the filtering and inverse Fourier transform, yields the desired amplitude \textbf{(e)} and phase \textbf{(f)} signals.}
		\label{fig:hologram}
	\end{figure}

Reconstructing the phase difference can be done via computational methods, selecting a side band in Fourier space and then transforming back, as shown in Figure~\ref{fig:hologram}. This process was partially automated using Holoworks 5 as part of the Gatan Digital Micrograph software suite. In all cases, after acquisition of the specimen hologram, an empty reference hologram was acquired with the same optical parameters, to allow for correction of any residual distortions or phase inhomogeneity arising from the optical system.   This process was used for reconstructing phase maps for all of the experimental data presented in this work.  Note that the primary objective lens, while powered down in Lorentz mode during all image and hologram acquisitions, could be temporarily excited in a controlled fashion to apply an in situ magnetic field in the specimen area, to achieve controlled magnetization of the specimen.
	
	\FloatBarrier
	\section{Characterization of Magnetic Needle as a Function of Temperature}
	
	In Figure 2 of the main text, we present results for holographic measurement 
	of the magnetic charge associated with the tip of a magnetic needle, extending 
	earlier work by Bech\'e {\it et al.} \cite{Beche2014}
	In modelling these results we, like Bech\'e {\it et al.}, assumed that 
	the phase ramp found in experiment originated from the magnetization of the needle [Figure 2c,d].
	Here we confirm that this is indeed the case, by showing that the phase ramp 
	in vacuum disappears, once the magnetic needle is heated above  
	its Curie point.

\begin{figure}[h]
		\subfloat[25$^\circ$C \label{fig:heating.25.celcius}]{
			\includegraphics[width=.25\textwidth]{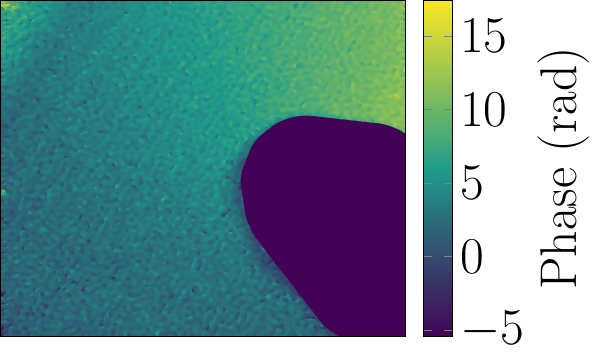}}
		\subfloat[75$^\circ$C]{
			\includegraphics[width=.25\textwidth]{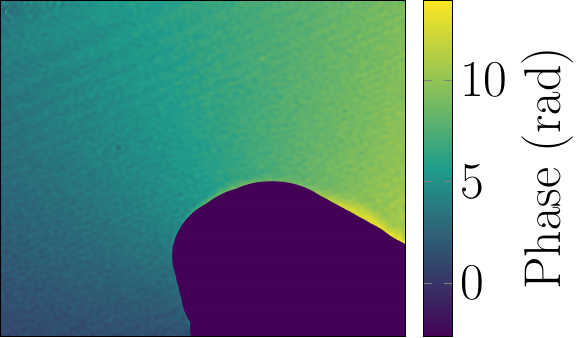}}
		\subfloat[125$^\circ$C]{
			\includegraphics[width=.25\textwidth]{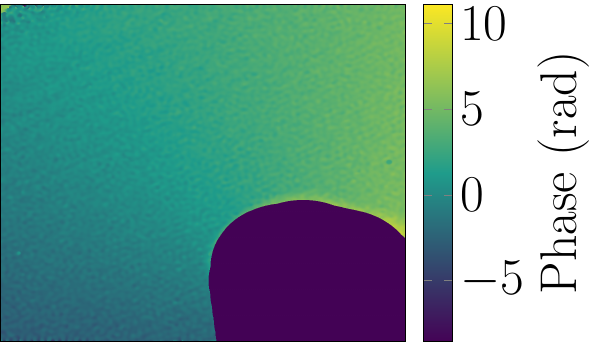}}
		\subfloat[175$^\circ$C]{
			\includegraphics[width=.25\textwidth]{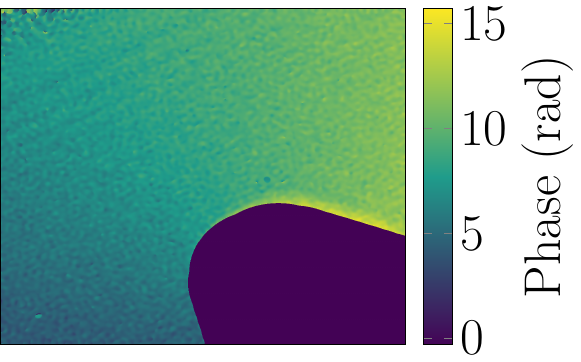}}\\
		\subfloat[225$^\circ$C]{
			\includegraphics[width=.25\textwidth]{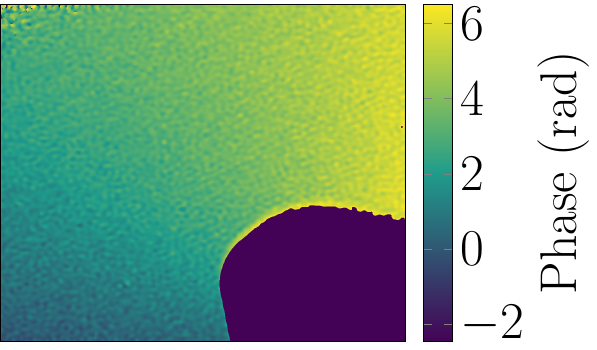}}
		\subfloat[275$^\circ$C]{
			\includegraphics[width=.25\textwidth]{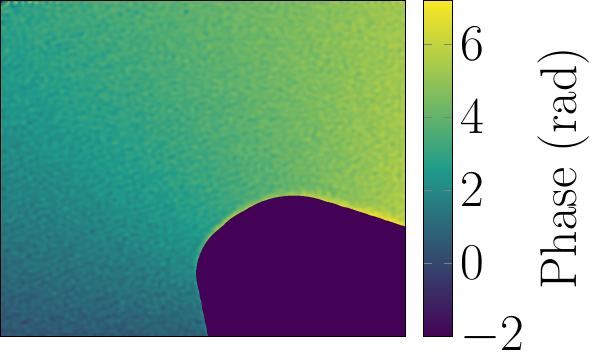}}
		\subfloat[325$^\circ$C \label{fig:heating.325.celcius}]{
			\includegraphics[width=.25\textwidth]{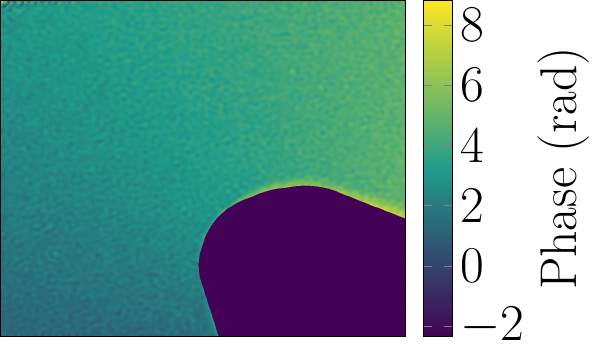}}
		\subfloat[375$^\circ$C \label{fig:heating.375.celcius}]{
			\includegraphics[width=.25\textwidth]{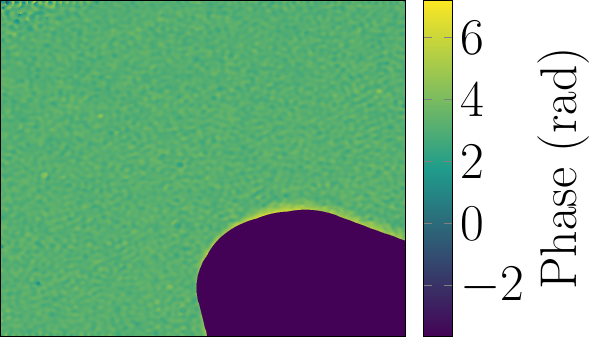}}\\
		\caption{\textbf{Phase maps of needle while heated past Curie point.} (\textbf{a-h}) Starting with a Ni needle magnetized along its long axis, phase shifts about the needle were recorded at regular temperature intervals while ramping up past the Curie point of 350$^\circ$C. The field of view is 150\,nm$\times$150\,nm. Note that the scale bar range reduces as temperature rises, since the phase shift is progressively weaker, until it wholly disappears at 375$^\circ$C. These results confirm the magnetization of the needle as the origin of the phase ramp described in the main text, and conclusively exclude any contribution from optical or numerical artifacts.}
		\label{fig:heating}
	\end{figure}

	In Figure~\ref{fig:heating} we show the phase maps reconstructed from a series of holographic measurements taken at temperatures ranging from 25$^\circ$C to 375$^\circ$C.  
	Heating was accomplished in--situ, using Gatan 626 TEM holder with heating element.
	Prior to heating, the magnetic domains within the needle were aligned by magnetizing it along the long axis, in situ, via manual control of the microscope's objective lens.
	A phase ramp is clearly visible in the initial measurements at 
	25$^\circ$C [Figure~\ref{fig:heating.25.celcius}, cf. Fig 2b of the main text]. 
	This phase ramp becomes progressively less pronounced on heating, 
	and is barely perceptible at 325$^\circ$C [Figure~\ref{fig:heating.325.celcius}].
	At 375$^\circ$C [Figure~\ref{fig:heating.375.celcius}], the needle has been heated 
	above the Curie point of Ni [$T_c \approx 365^\circ\text{C}$], and no phase ramp is observed.
	These results confirm the magnetization of the needle as the origin of the phase ramp described in the main text.

	\FloatBarrier
	\section{Artificial Spin Ice Characterization}
\begin{figure}[h!]
		\centering
		\includegraphics[height=.25\textheight]{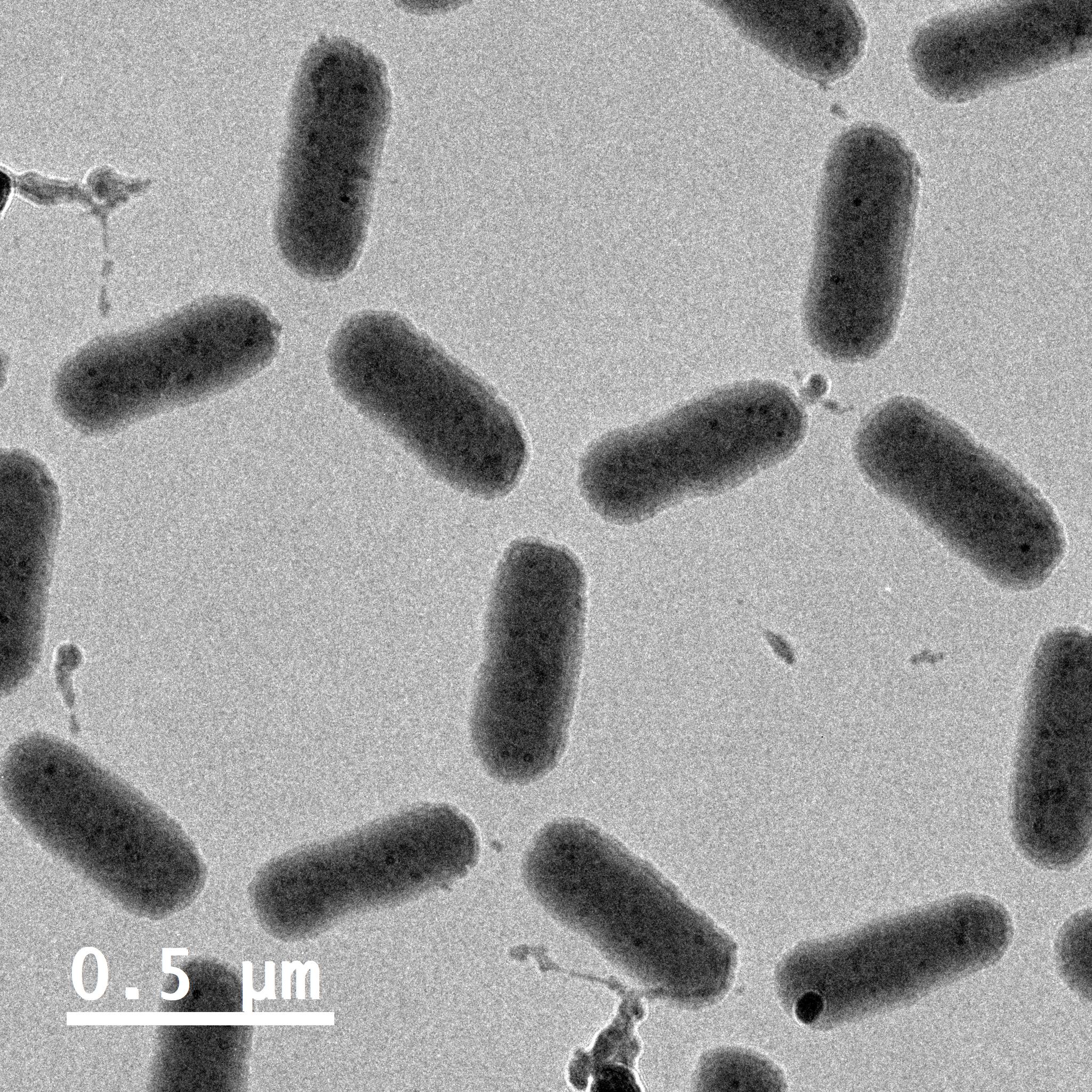}
		\caption{\textbf{In-focus image of artificial spin ice sample.} Scale bar is included to give a sense of the lattice spacing between adjacent vertices.}
		\label{fig:focus}
	\end{figure}

	In Figure 3 of the main text we present our results for studying artificial spin ice with electron holography. The results focus on vertices with a net charge of $3q$, where all three islands are pointing towards or away from the vertex center. In order to screen the artificial spin ice lattice for suitable vertices to image with holography, the general orientation of each magnetic island domain can be inferred through defocused Lorentz imaging [Figure 3a]. The contrast in this Fresnel mode of Lorentz imaging is affected by the electron beam being deflected perpendicular to the direction of the magnetic field in the sample.~\cite{Mankos1996323} Given the aspect ratio of the islands, they are expected to be remanently magnetized along the long axis, but the direction is unknown. The bright and dark fringes on either long side, arising from Lorentz deflection of the electrons, allow the respective magnetization directions to be inferred, thereby allowing $3q$ vertices to be identified.  Additionally, in-focus Lorentz imaging, as shown in Figure~\ref{fig:focus}, was utilized for accurately measuring the magnetic island dimensions and spacings.  The artificial spin ice lattice spacing was measured to be 580$\pm$11\,nm.

\begin{figure}[h!]
		\centering
		\includegraphics[width=.9\textwidth]{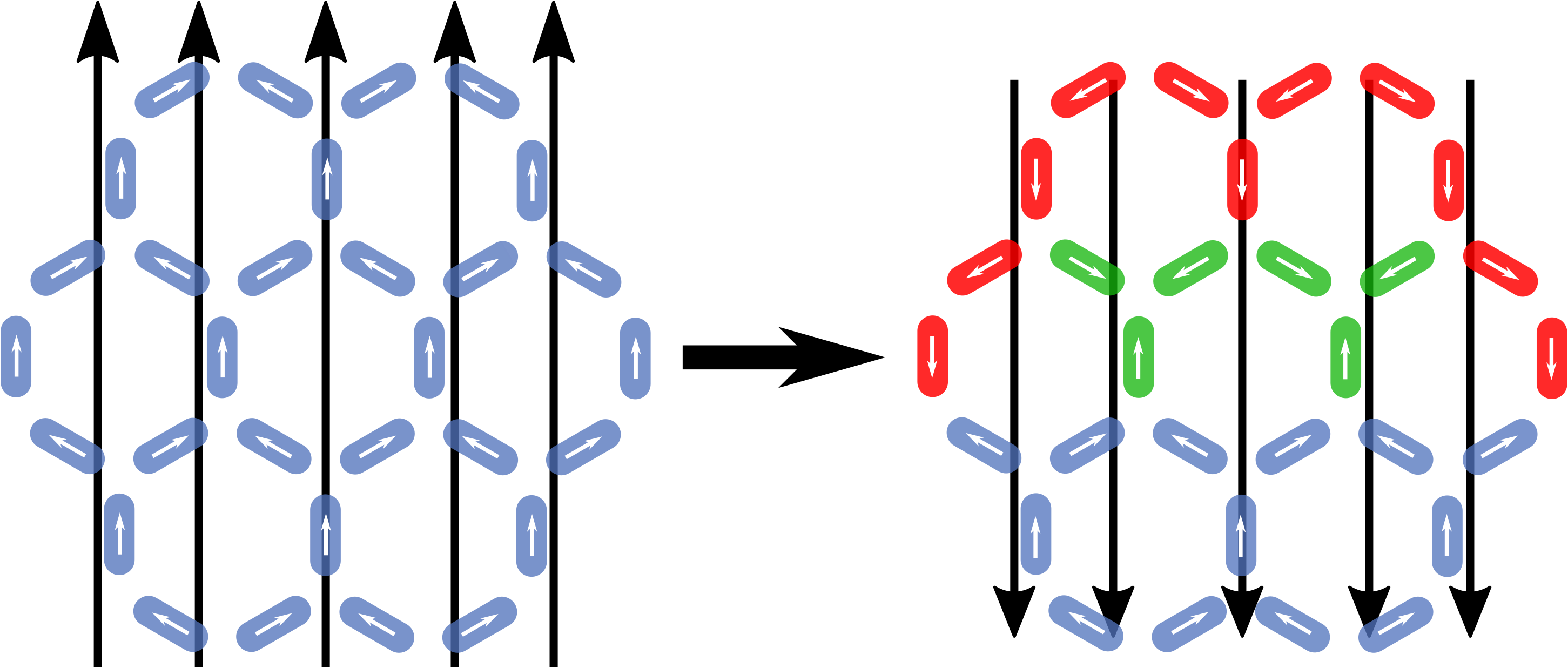}
	\caption{\textbf{Illustration of the methodology utilized to induce monopoles in artificial spin ice.} Applying a strong coercive field in one direction will force all islands along one direction (blue). By then applying a slightly weaker field in the opposite direction, islands will begin to flip in the opposite direction (red) causing monopoles to form at the boundary (green).  }
		\label{fig:ASIflip}
	\end{figure}

	To induce the formation of all-in/all-out ``3q'' vertices, an external field can be applied to the sample in situ, as illustrated in Figure~\ref{fig:ASIflip}.~\cite{Ladak2010,RealSpace} This is achieved by first tilting the sample 30$^\circ$, then applying the objective lens field to saturate the artificial spin ice along one direction, which requires about 5\% of the objective lens field (corresponding to 0.8kG flux density), and then applying a moderate amount (3\%) of the lens field in the opposite direction to partially reverse this saturation. This forms $3q$ monopoles along the boundary between oppositely magnetized regions, allowing them to be reproducibly generated. In addition, any vertices within these magnetized regions would have a slight phase ramp on top of their local phase signal, due to the correlated phase contribution from the surrounding islands.
	
	\FloatBarrier
	\subsection{Magnetic Moment Determination}
	The magnetic dipole moments of individual islands were determined from the experimental data. In-focus Lorentz micrographs were utilized to obtain the lateral dimensions, while reconstructed phase maps were utilized to obtain the permalloy layer thickness. The associated volumes, in conjunction with the known magnetization of permalloy (7.8$\times10^5$A/m),\cite{RealSpace} allow the magnetic dipole moments to be determined.  These measurements resulted in an average magnetic moment of $m = (2.9 \pm 0.3) \times 10^{-16} \text{A.m}^2$. The specific moments for individual islands were also used to match simulations to experimental phase maps.

	Note that the determination of permalloy layer thickness, as mentioned above, involved some careful treatment of the data. Firstly, the phase maps were corrected for any background phase distortions, utilizing reference holograms for standard correction, and additionally manual adjustment to correct any slight residual phase wedges. Secondly, the local phase shift arising from the electrostatic mean inner potential of permalloy (V$_{\rm 0}$ = 27.8 V)~\cite{kohn2005magnetic} was utilized to get a first estimate for permalloy thickness. Finally, the thickness value was then refined against the total experimental phase signal, in regions with and without permalloy. This comparison in thickness estimates from mean inner potential and generated phase signal for a given magnetic volume provided a sanity check on this measurement. This layer thickness value was then utilized for all subsequent magnetic dipole moment calculations.
	
	\section{Simulation Methods}
	Throughout the main text [Figures 2-5] we present simulations of electron holography on experimental magnetic monopole analogues at various length scales. Here we describe the finite-element simulation techniques used to generate those simulations.
	In order to simulate holography on magnetic systems, it is necessary to calculate the contribution to the electron phase from the magnetic vector potential (cf. Aharonov-Bohm effect). This requires calculating the magnetic vector potential $\vec{A}$ in the volume around each sample, then propagating the electron wave through it to sum up the net phase change $\Delta\phi$. Once this vector potential is calculated, any arbitrary electron beam path can be integrated through it to determine the effective phase shift. However, given that electron holography utilizes a wide, collimated, and coherent beam, it is reasonable to assume that the beam path is parallel and pointed entirely along the Z axis.~\cite{Beleggia2003} This means in reality that only the Z component of the vector potential is necessary to calculate the phase, reducing a path integral into the following one dimensional integral
	\begin{align}
		\Delta\phi_{AB}\bigg|_z &= -\frac{e}{\hbar}\oint \textbf{A}\cdot \textbf{dl}\bigg|_z\\
		&=  -\frac{e}{\hbar} \int A_z dz,\label{eq:ABint}
	\end{align}
	where $A_z$ is defined based on the system begin studied. Thus the phase map for a given sample can be calculated by integrating the vector potential around it along the Z axis. 
	
	In order to calculate this vector potential, it was necessary to first generate the arrangement of spins/magnetic domains for each experimental system. In the case of the needle and artificial spin ice [Figures 2c,3c] this can be done by breaking up the magnetic objects into sufficiently small magnetic domains ($\approxeq$5\,nm by 5\,nm by 5\.nm).  From there the vector potential for a given spin/domain in a given system is given by
	\begin{align}
		\textbf{A}(\textbf{r}_i)&=\frac{\mu_0}{4\pi}\sum_{j}\frac{\textbf{m}_j\times \textbf{r}_{i,j}}{|\textbf{r}_{i,j}|^3} \text{ in free space},\label{eq:vecpot}
	\end{align}
	where $m_i$ represents the magnetic moment for each spin/domain. Using these simulation methods the phase maps around various magnetic systems could be calculated to understand how holography would visualize spin phenomena.
	
	\subsection{Artificial Spin Ice}
\begin{figure}[h]
		\centering
		\includegraphics[width=0.25\linewidth]{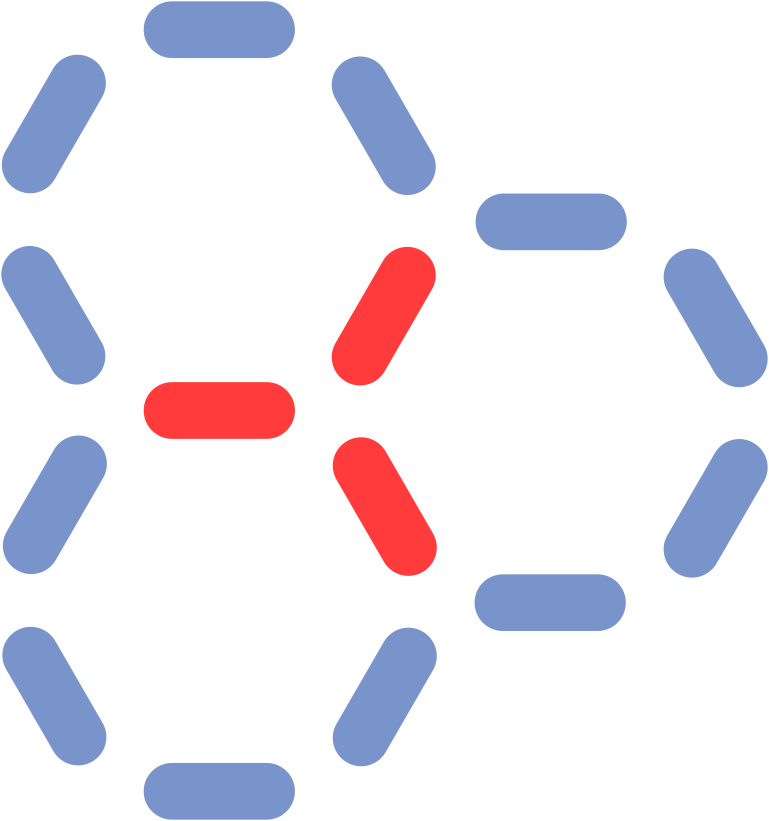}
		\caption{Overview of the surrounding islands (blue) that are simulated alongside the vertex of interest (red) to account for long range effects. These effects are minimal beyond this distance, so no further islands are required for simulating the details of one vertex.}
		\label{fig:ASIregion}
	\end{figure}

	These simulation methods were used to compare with experimental measurements of artificial spin ice monopole vertices [Figures 3c,e]. In order to ensure a good comparison between experiment and simulation, the size and shape of each simulated island was adjusted to match the experimental islands. For this reason the volume measurement of each island was helpful to ensuring this match. A given simulation would simulate the three islands of the vertex along with the 12 islands immediately connected to these three, as any further islands would not contribute strongly to the phase of the vertex. A representation of this region is shown in Figure~\ref{fig:ASIregion}.
	
	Another simplification that was taken to improve simulation performance was designing the simulation kernel to include the analytical form of Eq~\ref{eq:ABint} evaluated at $\pm\infty$, which would result in the following equation:
	\begin{align}
		-\frac{e}{\hbar} \int_{-\infty}^\infty \frac{\mu_0}{4\pi}\sum_{j,i}\frac{m_{yj}x_i-m_{xj}y_i}{(x_i^2+y_i^2+z_i^2)^{3/2}} dz = -\frac{e}{\hbar} \frac{\mu_0}{2\pi}T\sum_{j,i}\frac{m_{yj}x_i-m_{xj}y_i}{(x_i^2+y_i^2)} 
	\end{align}
	where $T$ is the thickness of the island, and the sum is applied over each point in the phase map ($i$), and the subdomains of each island ($j$). These simulated phase maps were then compared to experimental phase maps with both heatmaps [Figure 3b,c], as well as circular traces to ensure the phase map topology was correct, as shown below in Figure~\ref{fig:artificial.spin.ice.circ}.
	
	\begin{figure}[h]
		\centering
		\begin{tabular}{m{.249\textwidth}m{.249\textwidth}m{.249\textwidth}m{.249\textwidth}}
			\subfloat[]{\includegraphics[height=.15\textheight]{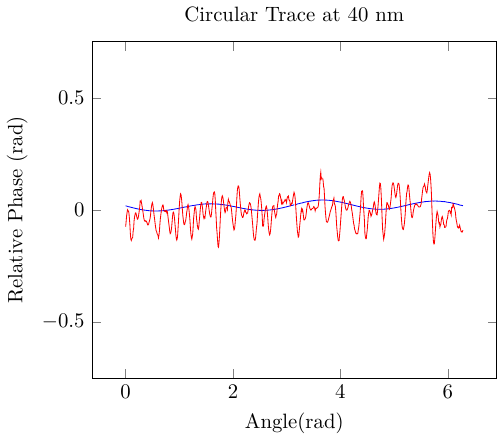}
				\label{fig:circular.trace.40nm}} & 
			\subfloat[]{\includegraphics[height=.15\textheight,clip=true,trim=15 0 0 0]{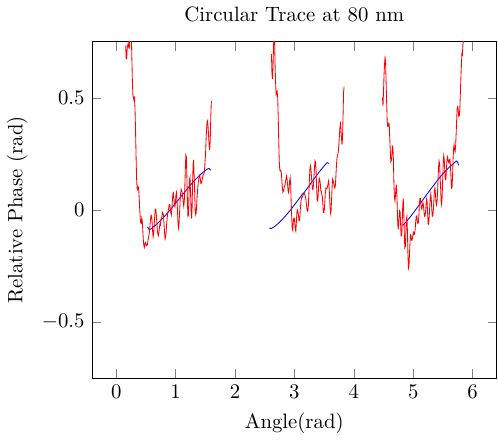}
				\label{fig:circular.trace.80nm}} &
			\subfloat[]{\includegraphics[height=.15\textheight,clip=true,trim=15 0 0 0]{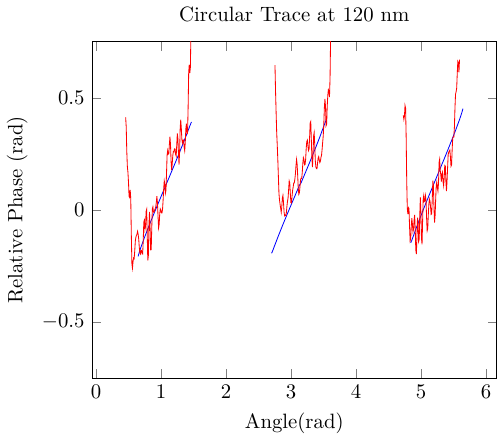}
				\label{fig:circular.trace.120nm}} &
			\subfloat[]{\includegraphics[height=.15\textheight,clip=true,trim=15 0 0 0]{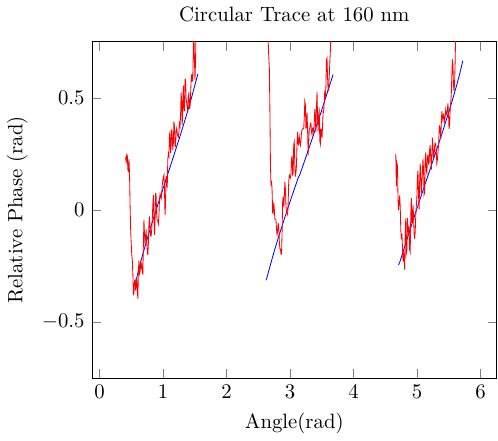}
				\label{fig:circular.trace.160nm}}
		\end{tabular}
		\caption{Circular trace around $3q$--vertex at radii of $40\ \text{nm}$, $80\ \text{nm}$, $120\ \text{nm}$, and $160\ \text{nm}$, showing the emergence of phase ramps between islands in both experiment and simulation (cf. Figure 3b,3c).  The data shown here is from the same sample as Figure 3 of the main text.
		}
		\label{fig:artificial.spin.ice.circ}
	\end{figure}

	\subsection{Pyrochlore Spin Ice}
	In Figures 4 and 5 of the main text we show simulations of how pyrochlore spin ice phase maps would appear, assuming a holography microscope with sufficient resolution were available. For the simulation of pyrochlore spin ice, the sample chosen was a thin film (1\,nm) of spin ice. This thickness corresponds with a single cubic cell of the pyrochlore lattice, meaning spins are distributed over 4 layers and form 3 layers of tetrahedra. To more accurately calculate a large lattice of magnetic spins and minimize boundary effects the sum is taken over a collections of spins in a periodic lattice. This means Eq~\ref{eq:vecpot} evolves to
	\begin{align}
		\textbf{A}(\textbf{r}_i)&=\frac{\mu_0}{4\pi}\sum_{j,n}\frac{\textbf{m}_j\times \left(\textbf{r}_{i,j}+L(n_{x}\hat{\textbf{x}}+n_{y}\hat{\textbf{y}})\right)}{|\textbf{r}_{i,j}+L(n_{x}\hat{\textbf{x}}+n_{y}\hat{\textbf{y}})|^3},
		\label{period}
	\end{align}
	The position vector is now modified by a lattice period $L=10$\,nm, which represents the size of a single unit of the periodic lattice. The number of periods in each direction $\textbf{n}$ was chosen based on how many would be required to keep the relative error between $n$ and $n+1$ periods less than 0.1\%. Based on this criterion 5 periods in each direction were used for the periodic sum for pyrochlore spin ice. 
	
\begin{figure}
		\centering
		\subfloat[]{\includegraphics[height=.155\textheight,clip=true,trim=0 0 70 0]{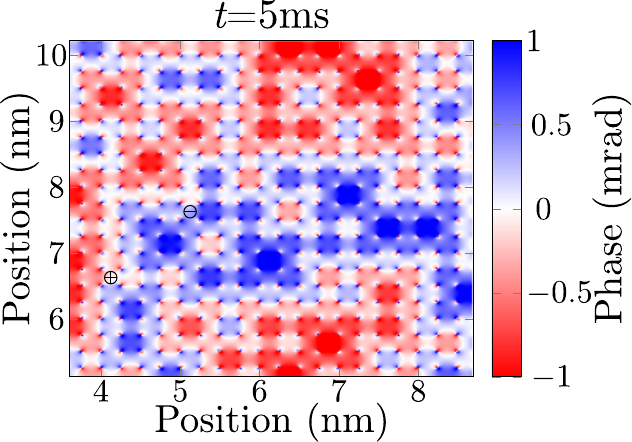}}
		\subfloat[]{\includegraphics[height=.155\textheight,clip=true,trim=0 0 70 0]{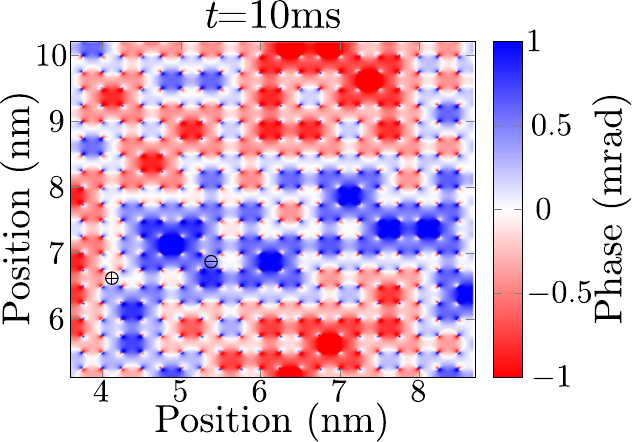}}
		\subfloat[]{\includegraphics[height=.155\textheight,clip=true,trim=0 0 70 0]{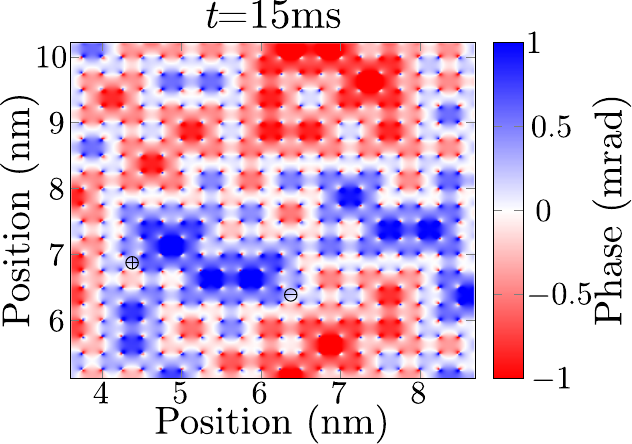}}
		\subfloat[]{\includegraphics[height=.155\textheight]{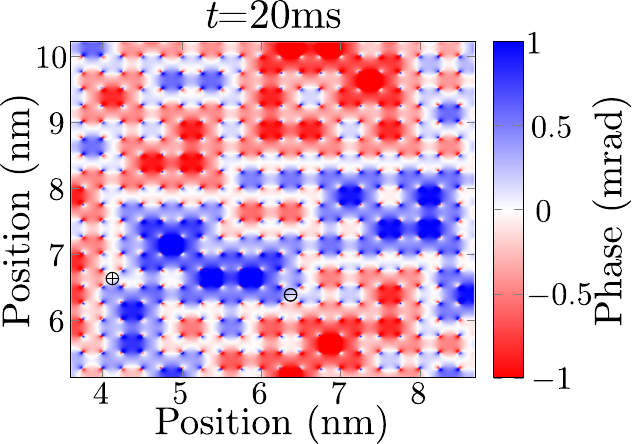}}\\
		\subfloat[]{\includegraphics[height=.155\textheight,clip=true,trim=0 0 70 0]{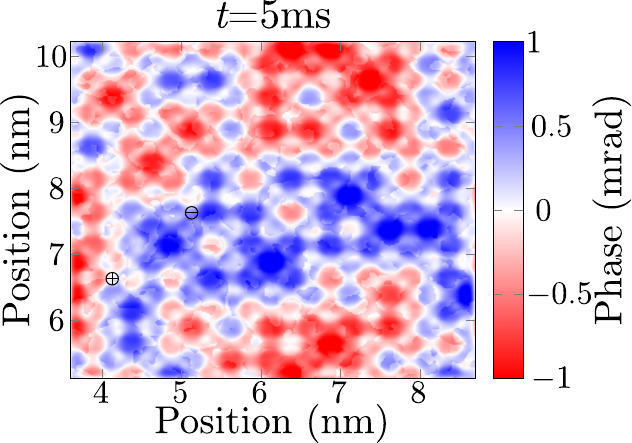}}
		\subfloat[]{\includegraphics[height=.155\textheight,clip=true,trim=0 0 70 0]{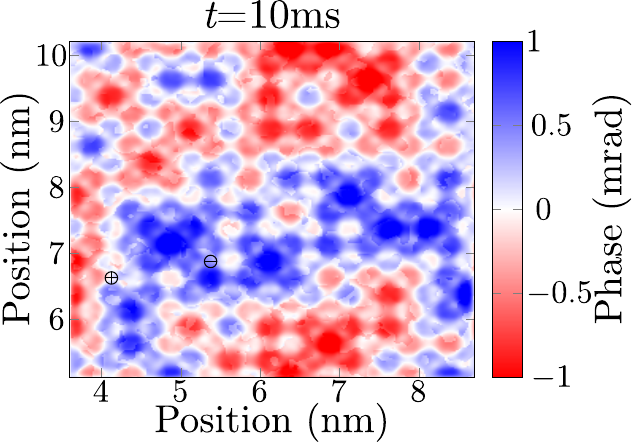}}
		\subfloat[]{\includegraphics[height=.155\textheight,clip=true,trim=0 0 70 0]{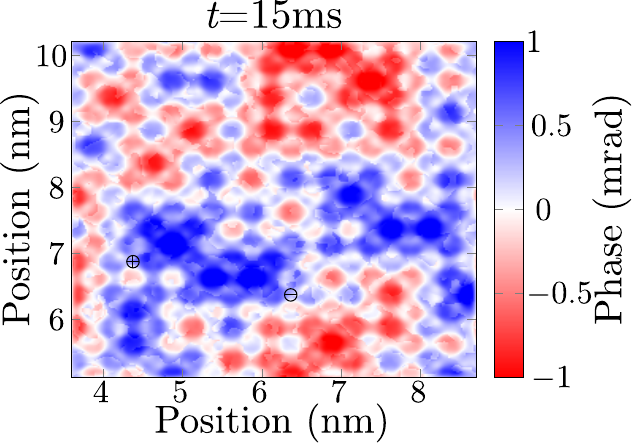}}
		\subfloat[]{\includegraphics[height=.155\textheight]{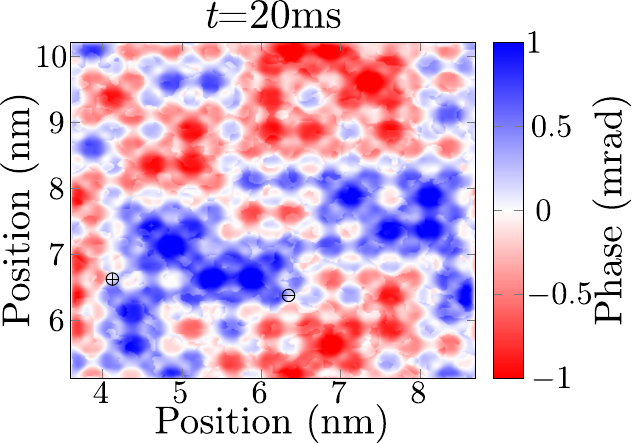}}\\
		
		\caption{\textbf{Simulated phase maps of a thermalized spin ice thin film at 700mK.} \textbf{(a-d)} These simulated phase maps at 20\,pm resolution, close to the theoretical ideal, show how monopoles (black) disrupt phase regions (red/blue) as they move through the lattice. \textbf{(e-h)} Once higher phase noise and limited spatial resolution is introduced however, this movement and disruption of phase regions becomes difficult to discern. This difficulty is the reason why using a differential measurement of the phase over time [Figure 5e-l] is preferable to visualize monopole movement.}
		\label{fig:thermalized}
	\end{figure}

	In practice this simulation was accomplished through the use of Fortran code written to sum the vector potential in a 3D volume around a sample. The volume is represented by a three-dimensional array where each element corresponds to one of the three components of magnetic vector potential ($A_x$,$A_y$,$A_z$), with spatial resolution of 20\,pm in each dimension. The simulation then goes through the volume via a loop, calculating the vector potential for a point in space based on all the contributions from the magnetic spins/domains in the sample and the periods in both X and Y directions. The location of each of these spins is read in from a file to easily test many different magnetic configurations. This summation is parallelized through MPI to allow for reasonable computation time. Once the vector potential around the thin film is calculated it is integrated into a phase map, as shown in Figure~\ref{fig:thermalized} as well as Figures 4a and 5a-d in the main text.
	
		\section{Animations of Monopole Dynamics in Pyrochlore Spin Ice}
	
	In Figure 5 of the main text we present a series of differential phase maps, showing how 
	dynamics of magnetic monopoles [Figure~\ref{fig:thermalizedN}] and electric fields 
	[Figure~\ref{fig:thermalizedE}] could be resolved through holographic measurements of 
	spin ice.
	These results were obtained through Monte Carlo simulations of the 
	time--evolution of spin configurations within a thin film of pyrochlore spin ice 
	at $700\ \text{mK}$, over a period of $1\ \text{s}$, sampled at an interval 
	of $2.5\ \text{ms}$, corresponding to a single Metropolis update of the entire lattice.
	For the purpose of Figure 5, this time series was further sampled at an interval 
	of $5\ \text{ms}$, over a period of $20\ \text{ms}$.
	Here we present phase maps for a $70\ \text{ms}$ segment of 
	the same time series in the form of two animations.
		
	In the first animation [movingMonopole.mp4], we show differential phase maps obtained 
	assuming spatial resolution of $20\ \text{pm}$ and perfect phase resolution, 
	equivalent to Figure 5e-h.
	Two monopoles are visible in the initial configuration at $t=0$.   
	Since the differential phase map is obtained by subtracting the phase 
	map at $t=0$ as a reference, it is entirely featureless at $t=0$.
	Over the next $\sim 15\ \text{ms}$ these monopoles diffuse independently around 
	the lattice, leading to a differential phase signal which tracks their motion.
	The charge associated with each monopole can now be identified from the 
	sense of the phase ramp it generates as it moves.
	And in the wake of the monopole motion, a Dirac string appears, connecting the 
	monopole with its initial position at $t=0$.

	A new element of spin--ice dynamics is visible from $t \approx 15\ \text{ms}$.
	Thermal fluctuations constantly create new pairs of monopoles.
	However these are short--lived, and typically annihilate, invisibly, 
	in less than $2.5\ \text{ms}$.
	However where monopoles traverse a closed loop before annihilating, 
	they leave a signal in the differential phase map, in the form of a region 
	of constant phase difference.
	And just as the movement of electric charge in a closed loop creates 
	a magnetic field, so the movement of magnetic charge in a closed loop 
	generates an electric field.~\cite{Hermele2004, Benton2012}
	
	The shortest closed loops on the pyrochlore lattice have the form of regular 
	hexagons in [111] planes (cf. Fig. 1 of the main text).
	In the [100] projection shown here, these present as elongated hexagons
	of constant phase difference.
	This phase difference can be positive or negative, depending on the 
	sign of the monopoles's charge, and sense in which it traversed the loop.
	In a thin--film geometry, it is also possible for monopoles to move on an open, 
	corkscrew path connecting the two edges of the sample.
	In the [100] projection, these present as unit--square regions of constant 
	phase difference.
	Both signals are visible in the animation.

	In the second animation [movingMonopoleNoise.mp4], we present exactly the same results, but 
	with simulated phase noise of $0.2\ \text{mrad}$, comparable to Figure 5f-l.
	In this case, the differential phase maps are subject to a constantly fluctuating 
	background.
	None the less, the signals coming from the movement of monopoles, and fluctuations 
	of electric field, are clearly visible.

	\section{Spatial Resolution and Phase Noise Estimates}
	In Figures 4 and 5 of the main text, we present phase maps of pyrochlore spin ice with ideal phase resolution, but also phase maps with finite spatial and phase resolution to estimate experimental limits for observation. In order to properly estimate the effects of lower spatial resolution and phase noise, spatial filters are applied to the high resolution phase maps. The reduced spatial resolution is generated by limiting the spatial information that can be reconstructed through holography. This is done by Fourier transforming the phase map, selecting only the information that represents length scales above the resolution limit via a circular aperture formed from a Gaussian window, and inverse transforming back into real space, represented as
	\begin{align}
		I_{filtered} = \mathcal{F}^{-1}\left[\mathcal{F}[e^{-i\phi_{ideal}}]\left(|\textbf{k}-\textbf{k}_{edge}|\leq \frac{2\pi}{\textbf{x}_{res}}\right)\right],\label{eq:filter}
	\end{align}
	where $\textbf{x}_{res}$ is the spatial resolution limit, and $\textbf{k}_{edge}$ represents the FWHM of the Gaussian window. The resultant image is then taken as the filtered phase map. The phase noise is generated as shot noise from a Poisson distribution and then run through the same spatial filtering before being added to the filtered phase map. The standard deviation of this resultant noisy phase map dictates the phase resolution, as with experimental phase maps. 

	\begin{figure*}[ht]
		\centering
		\begin{tabular}{p{.33\textwidth}p{.33\textwidth}p{.33\textwidth}}
			\subfloat[]{\includegraphics[height=.165\textheight]{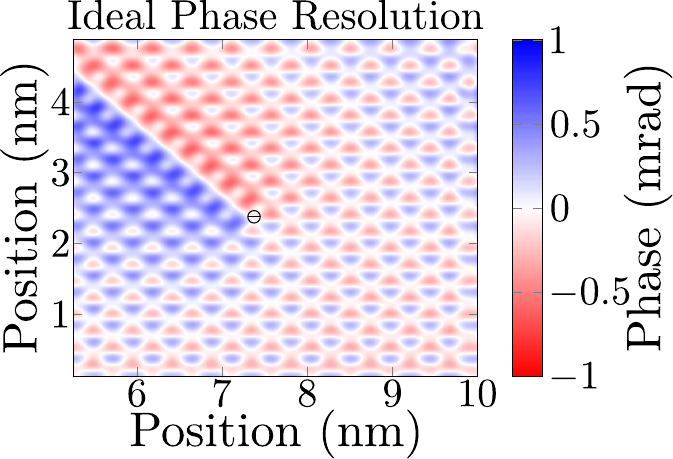}} &
			\subfloat[]{\includegraphics[height=.165\textheight]{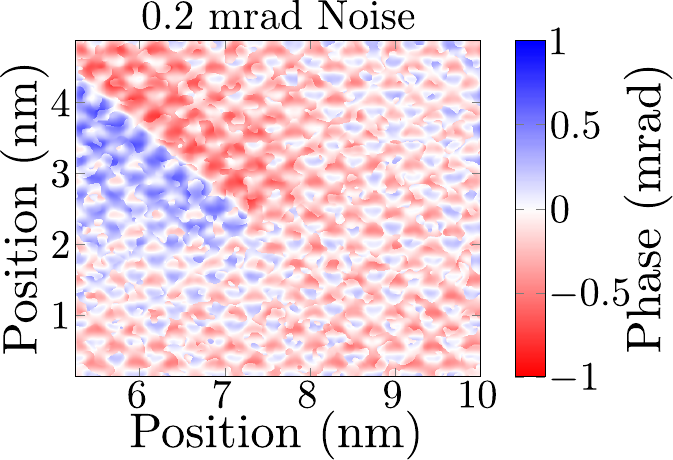}} &
			\subfloat[]{\includegraphics[height=.165\textheight]{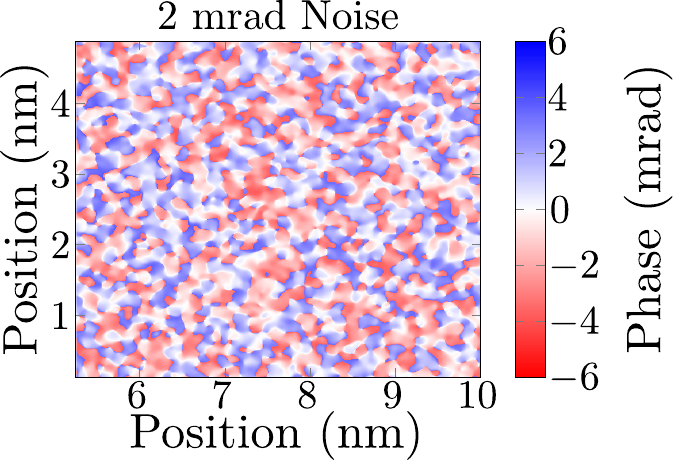}}\\
			\subfloat[]{\includegraphics[height=.2\textheight]{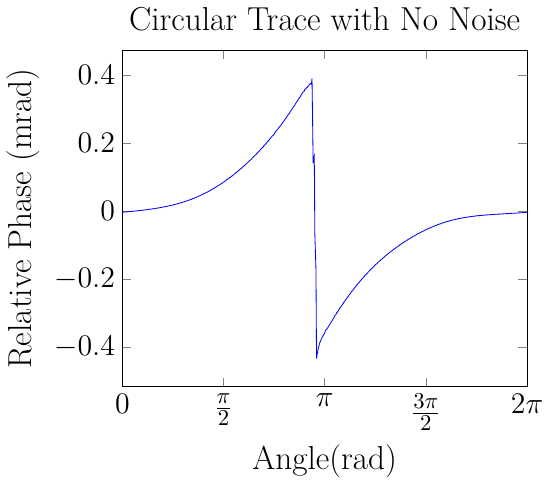}} &
			\subfloat[]{\includegraphics[height=.2\textheight]{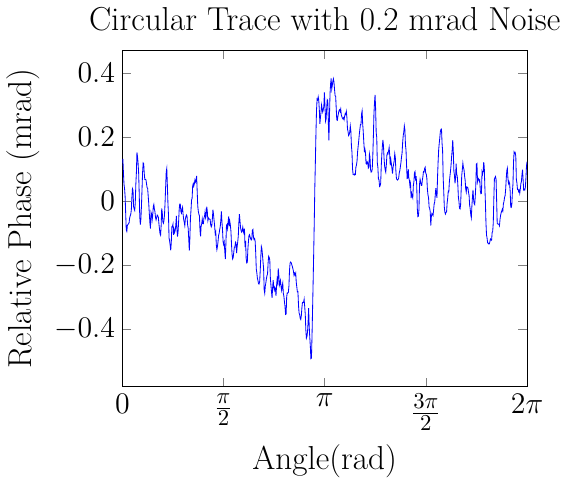}} &
			\subfloat[]{\includegraphics[height=.2\textheight]{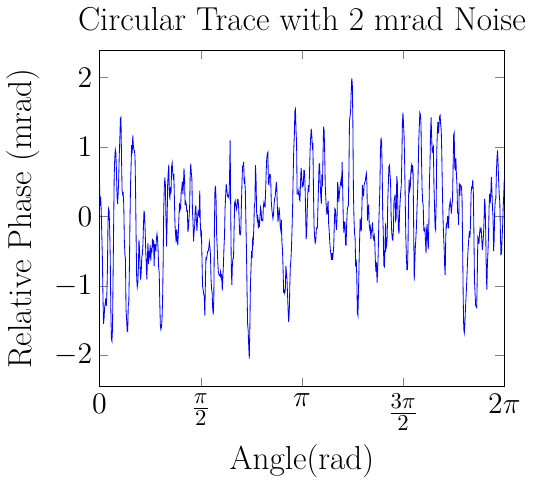}}
		\end{tabular}
		\caption{\textbf{Simulation of holographic measurement of magnetic monopole in a thin film of spin ice in applied magnetic field.} \textbf{(a)} Simulated phase maps with 240\,pm resolution give a sense of how the flipped spins reveal a Dirac string emerging from a single monopole ($\ominus$). The simulated thin film is 1\,nm thick with all spins magnetized along the Y axis. The Dirac string connected to each monopole breaks this order and divides the phase into distinct regions. A single -1 monopole ($\ominus$) draws a Dirac string with positive phase on the right and negative phase on the left. \textbf{(b,c)} Taking into account phase noise as well also shows this Dirac string can be easily identified at phase resolution levels of $0.2\ \text{mrad}$, but not quite at 2\,mrad. \textbf{(d-f)} Furthermore circular traces around each monopole show linear ramps similar to previous results. Each trace was taken at a radius of 2\,nm, with the background correction for phase contribution from the lattice. Although the sharp jump in phase across the Dirac string is initially distinguishable, it eventually becomes lost even at higher noise levels. These resolution limits of 240\,pm and 2\,mrad are an upper bound for the capabilities of Hitachi's Atomic Resolution Holography Microscope for field-free imaging, which is required for magnetic samples.~\cite{Tanigaki2019} The position axes are given in units of the cubic-unit-cell length $a = 1.00(1)$ nm for Dy$_2$Ti$_2$O$_7$. }
		\label{fig:singleMP}
	\end{figure*}
	
	Using this definition of phase resolution, it was possible to test the requirements needed to observe monopoles and similar signals. This was first applied to the test case of a single monopole in a magnetized lattice, as shown in Figure~\ref{fig:singleMP}. All of these phase maps were taken with a spatial resolution of 240\,pm, but with increasing amounts of phase noise. As with Figure 4 in the main text, a single monopole is well resolved at phase resolution of $0.2\ \text{mrad}$. However when scaling up to phase resolution levels of 2\,mrad, the monopole signal is completely washed out by noise for a single phase map.
	
\begin{figure}
		\centering
		\subfloat[]{\includegraphics[height=.165\textheight,clip=true,trim=0 0 85 0]{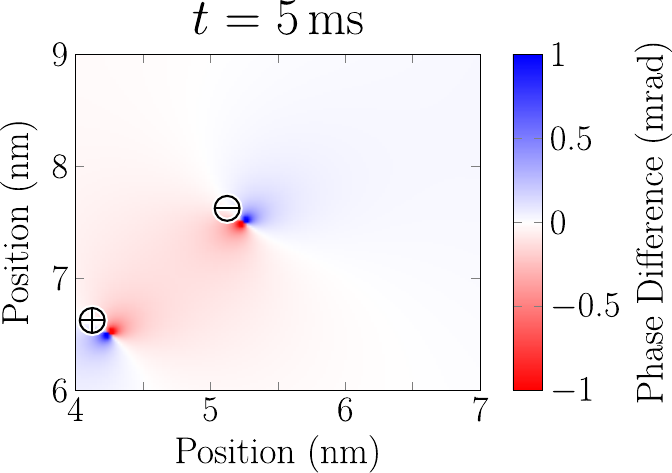}}
		\subfloat[]{\includegraphics[height=.165\textheight,clip=true,trim=0 0 85 0]{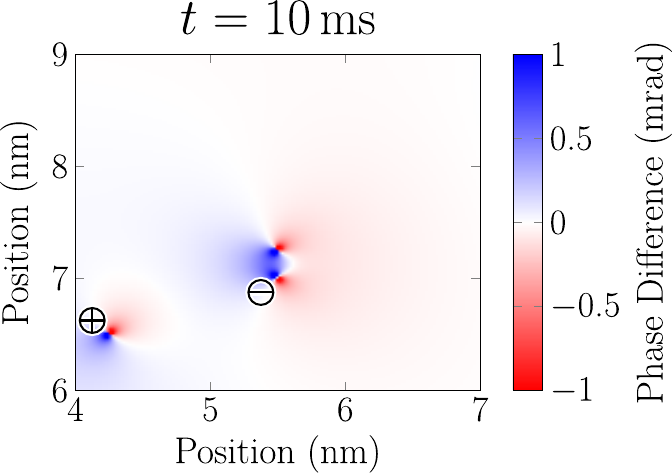}}
		\subfloat[]{\includegraphics[height=.165\textheight,clip=true,trim=0 0 85 0]{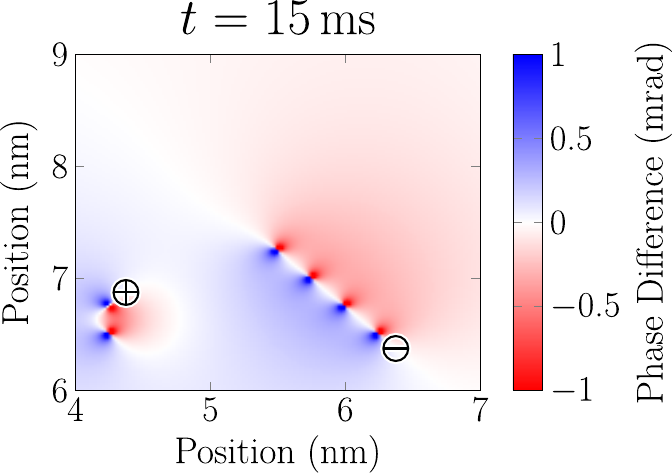}}
		\subfloat[]{\includegraphics[height=.165\textheight]{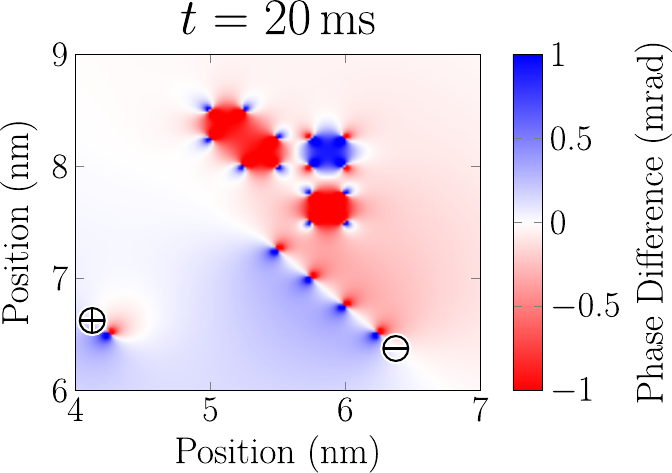}}\\
		\subfloat[]{\includegraphics[height=.165\textheight,clip=true,trim=0 0 85 0]{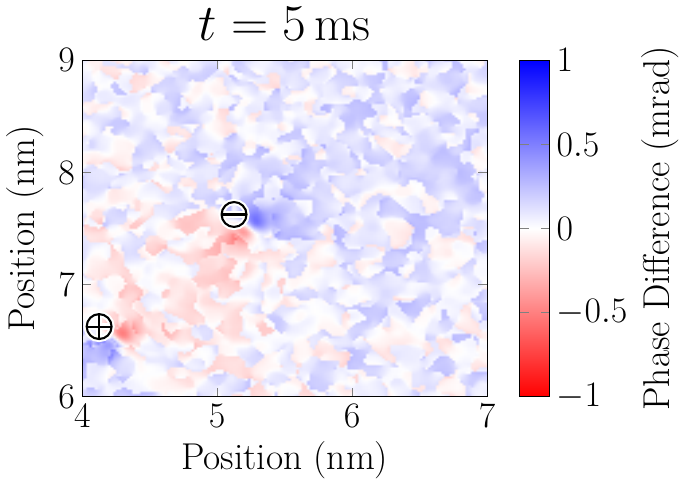}}
		\subfloat[]{\includegraphics[height=.165\textheight,clip=true,trim=0 0 85 0]{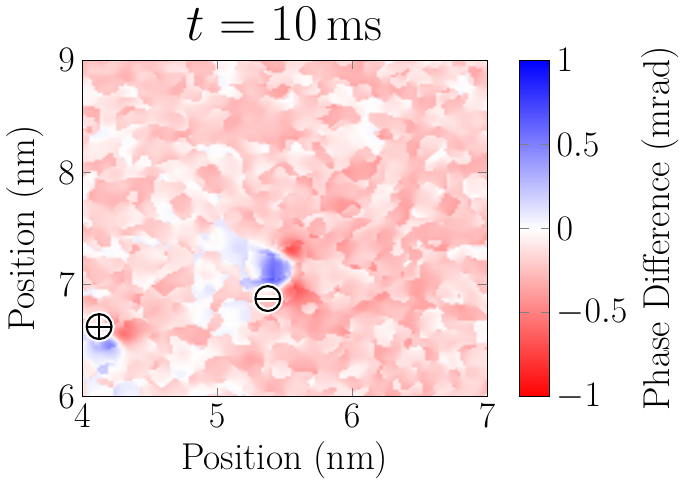}}
		\subfloat[]{\includegraphics[height=.165\textheight,clip=true,trim=0 0 85 0]{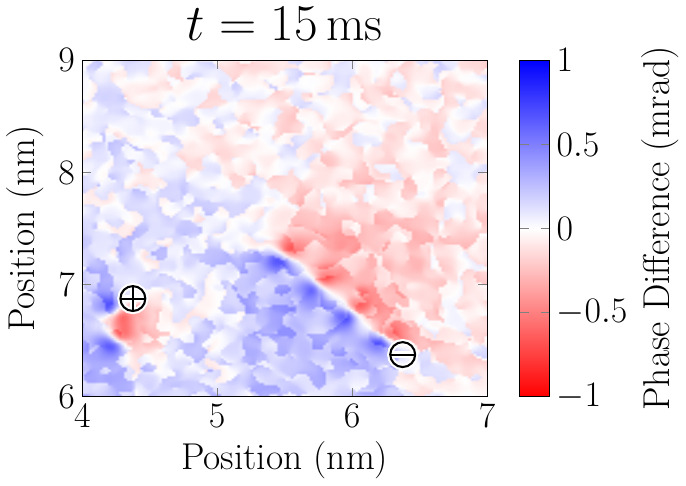}}
		\subfloat[]{\includegraphics[height=.165\textheight]{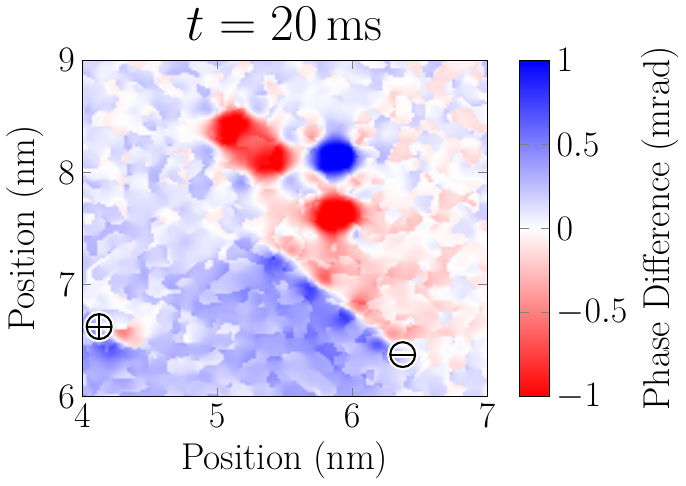}}\\
		\subfloat[]{\includegraphics[height=.165\textheight,clip=true,trim=0 0 85 0]{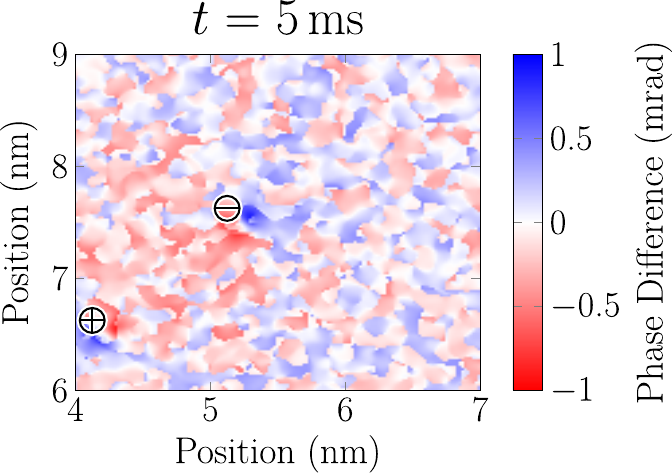}}
		\subfloat[]{\includegraphics[height=.165\textheight,clip=true,trim=0 0 85 0]{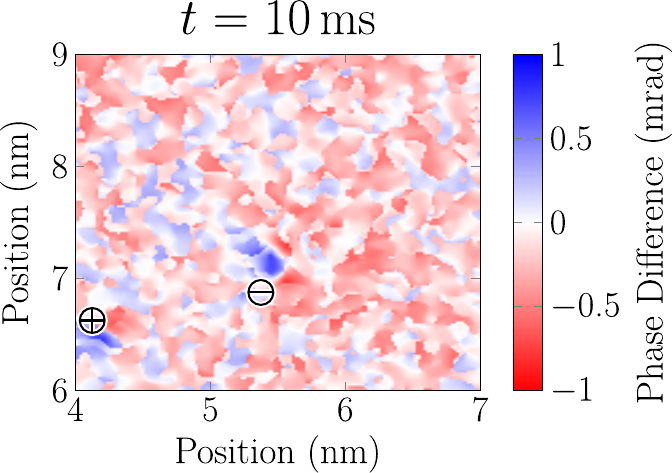}}
		\subfloat[]{\includegraphics[height=.165\textheight,clip=true,trim=0 0 85 0]{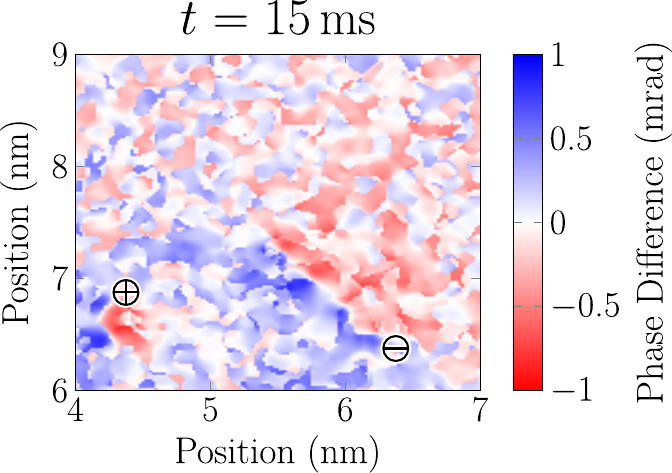}}
		\subfloat[]{\includegraphics[height=.165\textheight]{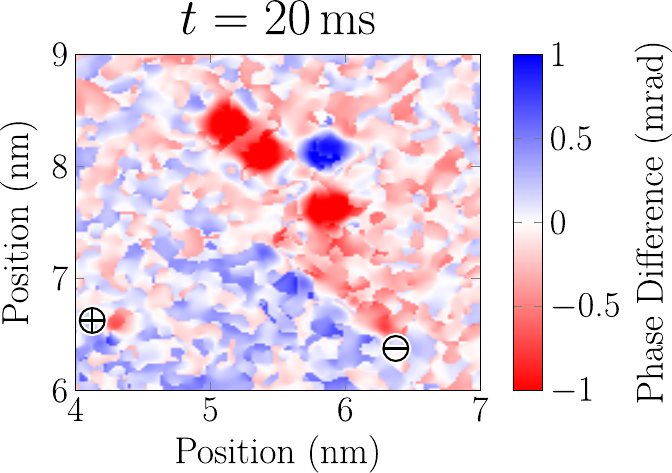}}\\	\subfloat[]{\includegraphics[height=.165\textheight,clip=true,trim=0 0 85 0]{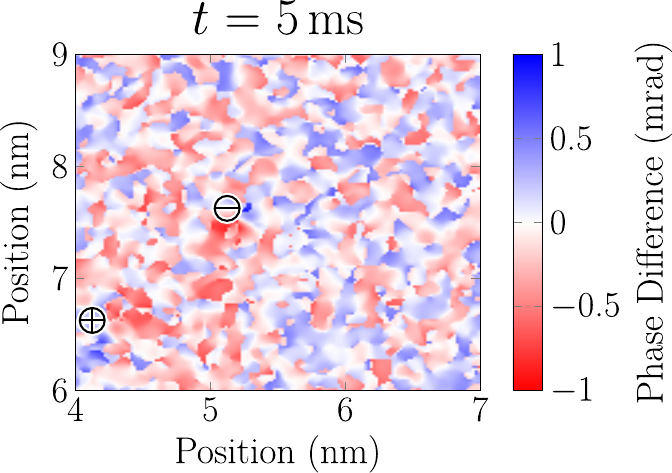}}
		\subfloat[]{\includegraphics[height=.165\textheight,clip=true,trim=0 0 85 0]{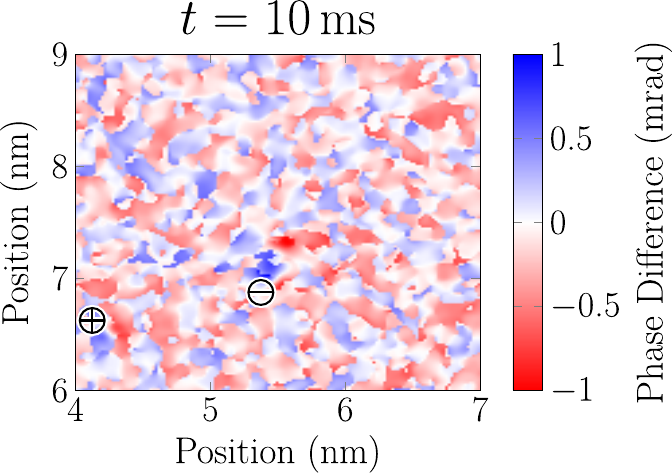}}
		\subfloat[]{\includegraphics[height=.165\textheight,clip=true,trim=0 0 85 0]{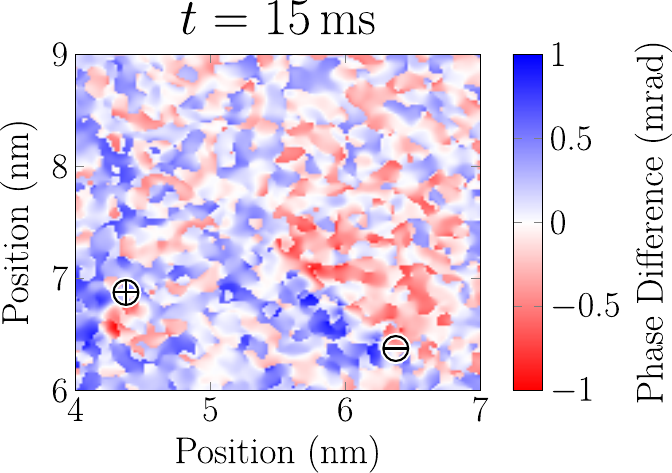}}
		\subfloat[]{\includegraphics[height=.165\textheight]{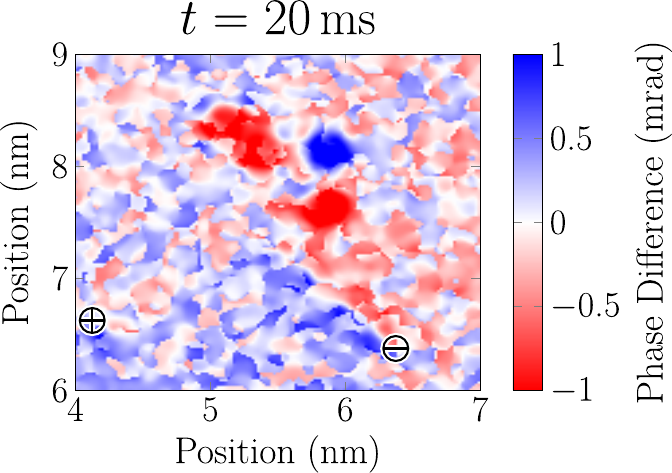}}\\	\subfloat[]{\includegraphics[height=.165\textheight,clip=true,trim=0 0 85 0]{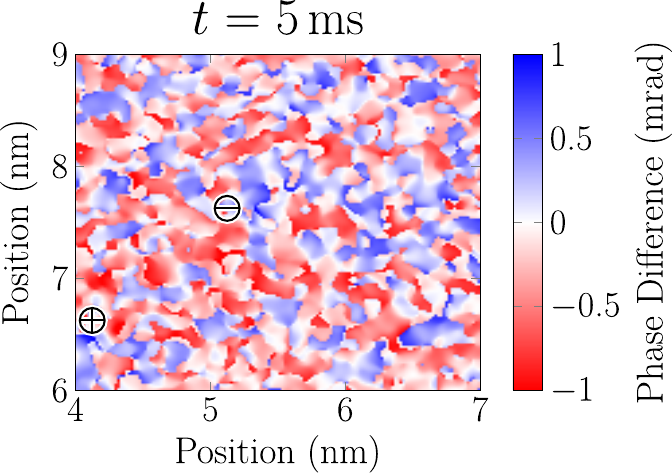}}
		\subfloat[]{\includegraphics[height=.165\textheight,clip=true,trim=0 0 85 0]{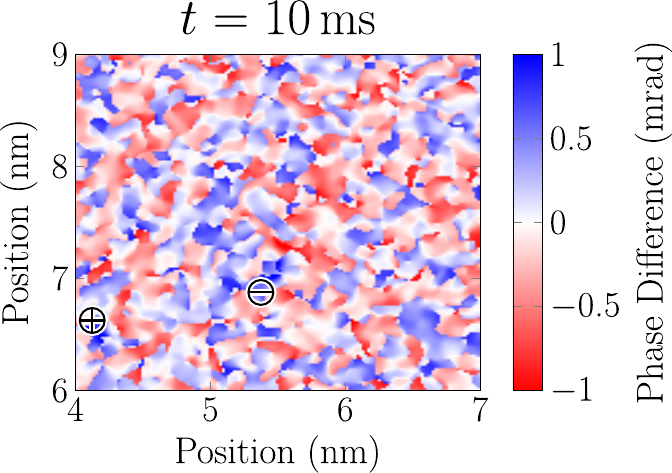}}
		\subfloat[]{\includegraphics[height=.165\textheight,clip=true,trim=0 0 85 0]{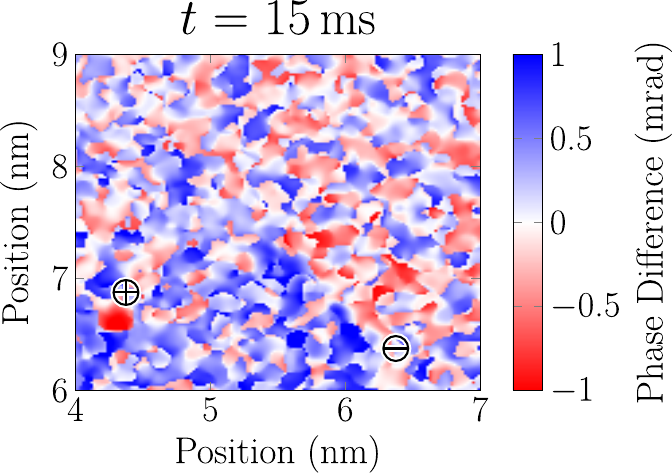}}
		\subfloat[]{\includegraphics[height=.165\textheight]{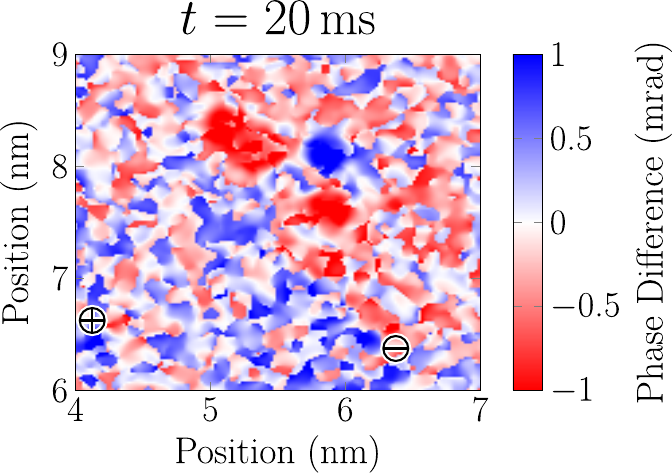}}\\
		\caption{\textbf{Simulated phase maps of a thermalized spin ice thin film at 700mK and increasing noise.} \textbf{(a-d)} These differential phase maps at 20\,pm resolution, close to the theoretical ideal, show how monopoles (black $\ominus$ and $\oplus$) trace differential signals in phase as they move within the lattice. \textbf{(e-h)} This movement remains distinguishable when considering phase noise at $0.1\ \text{mrad}$ or at  $0.2\ \text{mrad}$ \textbf{(i-l)}, but becomes harder to distinguish with $0.3\ \text{mrad}$ \textbf{(m-p)} and $0.4\ \text{mrad}$ noise \textbf{(q-t)}.}
		\label{fig:thermalizedN}
	\end{figure}

\begin{figure}
		\centering
		\subfloat[]{\includegraphics[height=.17\textheight]{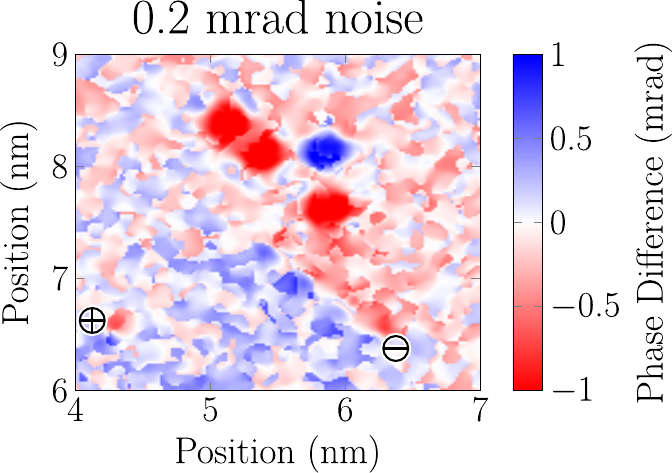}}
		\subfloat[]{\includegraphics[height=.17\textheight]{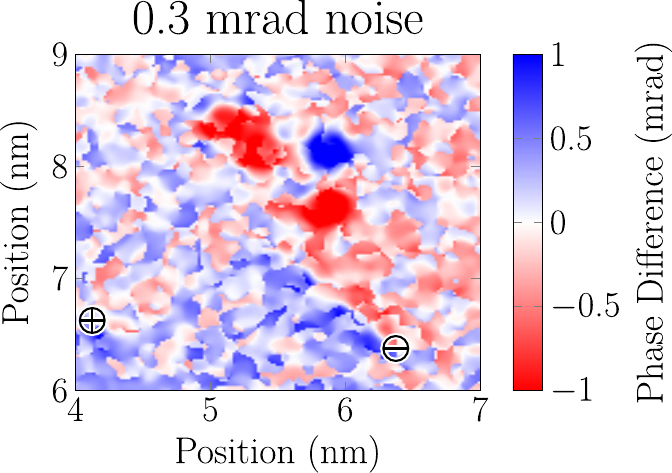}}
		\subfloat[]{\includegraphics[height=.17\textheight]{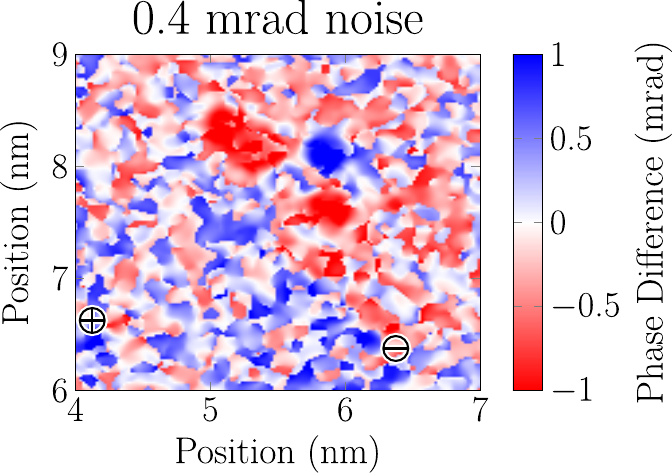}}\\
		\subfloat[]{\includegraphics[height=.17\textheight]{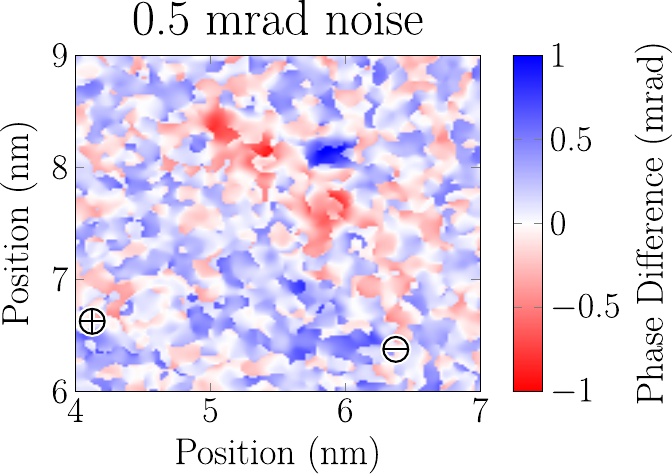}}
		\subfloat[]{\includegraphics[height=.17\textheight]{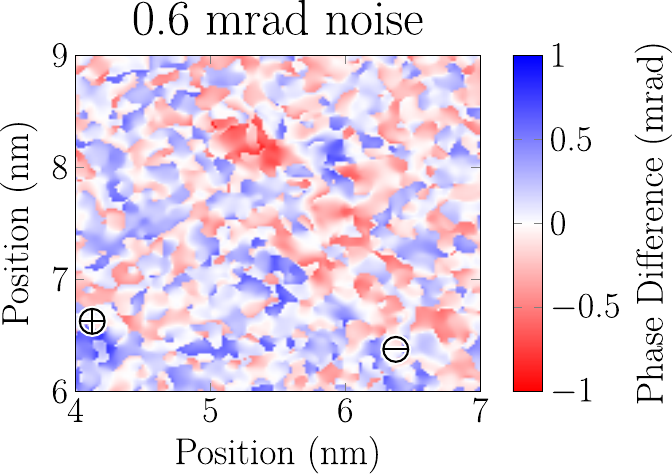}}
		\subfloat[]{\includegraphics[height=.17\textheight]{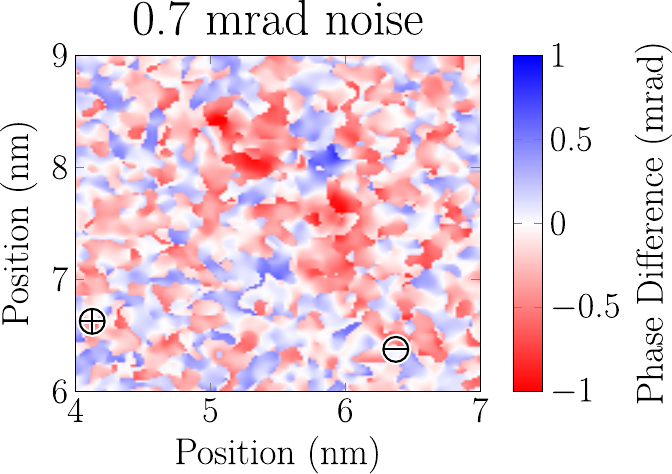}}\\
		\caption{\textbf{Simulated phase maps of a thermalized spin ice thin film at 700mK and emergent electric fields.} Alongside monopole movement, the emergence of electric field can be visualized with these same differential phase maps. Since these emergent fields are formed from small loops of flipped spins, their collective phase signature is stronger than a single monopole, allowing for their visualization at from 0.2 to 0.6\,mrad \textbf{(a-e)} before finally becoming indistinguishable from noise at 0.7\,mrad \textbf{(f)}}
		\label{fig:thermalizedE}
	\end{figure}

	Moving on to differential phase maps, a similar limit of $0.2\ \text{mrad}$ was set for resolving monopole movement through a thermalized lattice around 700\,mK [Figure 5i-l]. This is further visualized in Figure~\ref{fig:thermalizedN} where the same frames of monopole movement are simulated with increasing amounts of phase noise, from 0 to $0.4\ \text{mrad}$. Based on this comparison it becomes difficult to clearly identify monopole movement at $0.3\ \text{mrad}$, and completely impossible at $0.4\ \text{mrad}$. However, the electric field fluctuations in the final column of Figure~\ref{fig:thermalizedN} remain distinct even at these higher phase noise levels. Testing the limits of that particular signal in Figure~\ref{fig:thermalizedE} even further shows that these fluctuation can be resolved with a phase resolution of 0.6\,mrad, well above any monopole signals. This suggests that the electric field fluctuations also provide an important target for future holography studies, as equipment and specifications improve.
		
	\FloatBarrier

	\bibliographystyle{apsrev4-1}
	\bibliography{Supplemental}